\documentclass[jgrga,draft]{agutex}
\usepackage{graphicx}
\usepackage{tabularx}
\usepackage{lineno}
\usepackage{epstopdf}
\usepackage{rotating}

\setkeys{Gin}{draft=false}
\authorrunninghead{YEO ET AL.}
\titlerunninghead{UV SSI in NRLSSI and SATIRE-S}
\authoraddr{Corresponding author: K.~L.~Yeo, Max-Planck-Institut f\"{u}r Sonnensystemforschung, Justus-von-Liebig-Weg 3, 37077 G\"{o}ttingen, Germany. (yeo@mps.mpg.de)}

\begin{document}
\title{UV solar irradiance in observations and the NRLSSI and SATIRE-S models}
\author{K.~L.~Yeo,\altaffilmark{1} W.~T.~Ball,\altaffilmark{2} N.~A.~Krivova,\altaffilmark{1} S.~K.~Solanki,\altaffilmark{1,3} Y.~C.~Unruh\altaffilmark{4}, and J.~Morrill\altaffilmark{5}}
\altaffiltext{1}{Max-Planck-Institut f\"{u}r Sonnensystemforschung, G\"{o}ttingen, Germany.}
\altaffiltext{2}{Physikalisch-Meteorologisches Observatorium Davos, World Radiation Center, Davos, Switzerland.}
\altaffiltext{3}{School of Space Research, Kyung Hee University, Yongin, Gyeonggi, Korea.}
\altaffiltext{4}{Astrophysics Group, Blackett Laboratory, Imperial College London, London, United Kingdom.}
\altaffiltext{5}{Naval Research Laboratory, Washington DC, USA.}
\begin{abstract}
Total solar irradiance and UV spectral solar irradiance have been monitored since 1978 through a succession of space missions. This is accompanied by the development of models aimed at replicating solar irradiance by relating the variability to solar magnetic activity. The NRLSSI and SATIRE-S models provide the most comprehensive reconstructions of total and spectral solar irradiance over the period of satellite observation currently available. There is persistent controversy between the various measurements and models in terms of the wavelength dependence of the variation over the solar cycle, with repercussions on our understanding of the influence of UV solar irradiance variability on the stratosphere. We review the measurement and modelling of UV solar irradiance variability over the period of satellite observation. The SATIRE-S reconstruction is consistent with spectral solar irradiance observations where they are reliable. It is also supported by an independent, empirical reconstruction of UV spectral solar irradiance based on UARS/SUSIM measurements from an earlier study. The weaker solar cycle variability produced by NRLSSI between 300 and 400 nm is not evident in any available record. We show that although the method employed to construct NRLSSI is principally sound, reconstructed solar cycle variability is detrimentally affected by the uncertainty in the SSI observations it draws upon in the derivation. {Based on our findings, we recommend, when choosing between the two models, the use of SATIRE-S for climate studies.}
\end{abstract}

\begin{article}

\section{Introduction}
\label{introduction}

The assessment of human impact on the Earth's climate requires an understanding of the natural drivers of climate change. This includes the variations in the solar radiative flux impinging on the Earth system, usually described in terms of solar irradiance. There is growing evidence of its influence on the Earth's climate \citep[see][and references therein]{haigh07,gray10,solanki13}. As solar irradiance variability and its interaction with the Earth's atmosphere is wavelength dependent, accurate knowledge of both total and spectral solar irradiance (TSI and SSI) variability is pertinent to establishing the solar contribution to climate change.

{TSI has been tracked from space, largely without interruption, since 1978 through a succession of satellite missions  \citep[see recent reviews by][]{frohlich12,kopp14}. Similarly, ultraviolet (UV) SSI has been monitored by a medley of spaceborne instruments covering various periods and wavelengths over the same period \citep[see][and Sect. \ref{review}]{deland12,ermolli13}. Except below 170 nm before 1992, the various satellite records collectively represent nearly uninterrupted daily sampling of SSI between 120 and 400 nm since 1978 (Figs. \ref{coverage} and \ref{absolute}).} Monitoring of the visible and infrared started much later, in 2003, with the SIM instrument onboard SORCE \citep[covering 200 to 2416 nm,][]{harder05a,harder05b,rottman05}. (The definition of the abbreviations used in this paper, omitted in the text at parts to ease readability, is summarized in Table \ref{acronyms}.)

{The interpretation of the body of satellite measurements is hampered by differences in the absolute radiometry and the overall trend with time between the various records. These discrepancies arise from uncertainties in the instrument calibration and the account of effects such as exposure/time degradation and sensitivity changes.} {Even so, the} body of satellite observations {has revealed} correlations of solar irradiance variability with the passage of bright/dark magnetic structures across the solar disc \citep{willson81,hudson82} and with the 11-year solar magnetic cycle \citep{willson88,hickey88}. This led to the development of models aimed at reconstructing solar irradiance by relating the variability at timescales greater than a day to solar magnetic activity \citep{oster82,foukal86}. At shorter timescales, contributions from acoustic oscillations, convection and flares become significant \citep{hudson88,woods06,seleznyov11}. Solar irradiance variability is modelled as the influence of sunspots and pores and of the magnetic concentrations that make up active region faculae and quiet Sun network, usually termed sunspot darkening and facular brightening, respectively. This information is inferred from solar observations indicative of magnetic activity \citep[recently reviewed by][]{ermolli14}.

\cite{ermolli13} and \cite{yeo14a} reviewed the present-day models of solar irradiance capable of returning TSI and SSI simultaneously. They are, NRLSSI \citep{lean97,lean00}, SATIRE \citep{unruh99,fligge00,krivova03,krivova09,wenzler06,ball11,ball12,ball14a,yeo14b}, SRPM \citep{fontenla99,fontenla04,fontenla06,fontenla09,fontenla11}, OAR \citep{penza03,ermolli03,ermolli11,ermolli13} and COSIR \citep{haberreiter05,haberreiter08,shapiro10,shapiro11,shapiro13,cessateur15}. The approach taken by these models is to reconstruct the solar spectrum over a wavelength range stretching at least from 120 nm to the far IR such that the integral over wavelength is a good approximation of TSI. Solar radiative flux shortwards of 120 nm \citep{lilensten08,woods08} and in the radio \citep{tapping13,dudokdewit14} constitute a minute proportion of the total flux and while the variability affects space weather \citep[e.g.,][]{schwenn06}, it exerts little influence on the Earth's climate. By giving both TSI and SSI over the wavelength range pertinent to Sun-climate interactions, such models are particularly relevant to climate studies. The models listed make use of various solar observations to reconstruct solar irradiance over different periods. Currently, only NRLSSI and SATIRE provide reconstructions of TSI and SSI encompassing the entire period of satellite observation at daily cadence. SATIRE-S, the version of SATIRE based on full-disc magnetograms and intensity images, was recently updated and extended to the present by \cite{yeo14b}.

{Apart from the models which reconstruct both TSI and SSI simultaneously, there are also models with the more limited scope of returning either TSI \citep[such as,][]{chapman13,frohlich13} or UV SSI \citep{morrill11,thuillier12,bolduc12} alone.}

Models of solar irradiance have been successful in replicating most of the apparent variability in TSI and in certain SSI records \citep[recent examples include][]{bolduc12,chapman13,yeo14b}, confirming the dominant role of solar magnetic activity in driving solar irradiance variability over the period of satellite observation \citep{domingo09,yeo14a}. While significant progress has been made in the measurement and modelling of solar irradiance variability, the secular trend over the satellite era \citep{scafetta09,krivova09,frohlich12,yeo14b,ball14b} and critically for climate studies, the wavelength dependence of the variation over the solar cycle remain controversial \citep[][]{ball11,deland12,lean12,unruh12,ermolli13,solanki13,wehrli13,morrill14,yeo14a,yeo14b}.

{SORCE/SIM} (version 19, dated November 2013) registered a drop in 240 to 400 nm SSI between 2003 and 2008, within the declining phase of solar cycle 23, that is about ten and five times greater than what is replicated by NRLSSI and SATIRE-S, respectively \citep[see Fig. 7 in][]{yeo14a}. This decline also exceeds projections from pre-SORCE measurements by similar margins. Studies have also noted a similar disagreement between the SSI measurements from SORCE/SOLSTICE \citep[up to version 12, dated September 2012, covering 115 to 310 nm,][]{mcclintock05,snow05a} and both models and pre-SORCE measurements above 180 nm. The disparity is, however, far less acute than what is seen with {SORCE/SIM}. With the discrepant solar cycle variability longwards of 180 nm, the application of NRLSSI, SATIRE-S and SORCE SSI to climate models yield qualitatively different results for the response in stratospheric ozone, temperature and heating rates \citep{haigh10,merkel11,oberlander12,ermolli13,dhomse13,ball14a,ball14b}. {Revisions of the SORCE/SIM (versions 20 to 22) and SORCE/SOLSTICE (versions 13 and 14) records have since been released, with the latest made available in May 2015.}

The focus and aim of this paper is to assess the UV SSI variability produced by NRLSSI and by SATIRE-S, and therefore the applicability of these models to climate studies. In Sect. \ref{models}, we describe the NRLSSI and SATIRE-S reconstructions. We will also introduce the empirical reconstruction of UV SSI by \citealt{morrill11} (denoted Mea11), which we will match against the two other reconstructions in the subsequent discussion. We then examine the wavelength dependence of solar irradiance variability in the UV in the body of satellite observations{, including the recent revision of the SORCE/SIM and SORCE/SOLSTICE records,} and the various reconstructions (Sect. \ref{ssi}). This is followed by a summary and concluding statements in Sect. \ref{conclusions}.

\section{Models of Solar Irradiance Examined in the Present Study}
\label{models}

\subsection{Naval Research Laboratory Solar Spectral Irradiance (NRLSSI)}

In the NRLSSI model, the time variation of sunspot darkening is given by the photometric sunspot index, PSI \citep{frohlich94,lean98} and that of facular brightening by a combination of the NOAA Mg II index composite \citep{viereck04} and the Mg II index record based on SORCE/SOLSTICE SSI \citep{snow05b}.

The reconstruction spans 120 to 100000 nm {(0.1 mm)}. For the 120 to 400 nm segment, solar irradiance variability is determined from the multiple linear regression of the index data to the SSI record from UARS/SOLSTICE \citep{rottman01}. Longwards of around 220 nm, solar cycle variability is obscured in the UARS/SOLSTICE record by the long-term uncertainty \citep{woods96}. To factor this out of the analysis, the regression is limited to the variability over solar rotation timescales by first detrending the index and SSI time series. Solar irradiance variability is then recovered by applying the regression coefficients to the original (i.e., not detrended) index data. In doing so, the model assumes solar irradiance scales linearly with the various indices and the apparent relationship between the two at solar {rotation} timescales applies at solar {cycle} timescales. Solar irradiance variability above 400 nm is given by modulating the intensity contrast of sunspots and faculae in time by the index data. The sunspot and facular intensity contrasts were taken from \cite{solanki98}. {The two intensity contrasts data sets} are scaled in NRLSSI such that the integrated variability over the wavelength range of the model matches that from TSI modelling \citep{frohlich04}{, introducing two free parameters into the model.} Finally, the reconstructed solar irradiance variability is imposed on an intensity spectrum of the quiet Sun composed from UARS/SOLSTICE and SOLSPEC/ATLAS-1 \citep{thuillier98} measurements and the \cite{kurucz91} calculations.

\subsection{Spectral And Total Irradiance REconstruction for the Satellite era (SATIRE-S)}
\label{modelssatires}

SATIRE-S \citep{fligge00,krivova03,krivova09} is what is termed a semi-empirical model, along with SRPM, OAR and COSIR. Semi-empirical models take a somewhat less empirical and more physics-based approach than models based on relating indices of solar magnetic activity to measured solar irradiance, relying instead on the intensity spectra of solar surface features generated with radiative transfer codes to recreate the solar spectrum.

In SATIRE-S, the solar surface is described as comprising of quiet Sun, faculae and sunspots. The surface coverage of these components is determined from full-disc magnetograms and intensity images {taken close together in time}. The recent update of the model by \cite{yeo14b} employed such observations from the Kitt Peak vacuum telescope \citep{livingston76,jones92}, SoHO/MDI \citep{scherrer95} and SDO/HMI \citep{schou12} to reconstruct TSI and SSI from 115 nm to 160000 nm {(0.16 mm)} back to 1974. {Apart from extending the preceding version of the model \citep[spanning 1974 to 2009,][]{ball14b}, which was based on similar Kitt Peak and SoHO/MDI observations, to the present with SDO/HMI data, the reconstruction method was also modified. We will describe the model in its present form, as detailed in \cite{yeo14b}, before discussing these modifications.} The Kitt Peak, {SoHO/MDI} and {SDO/HMI magnetogram} data sets were {corrected for differences in instrument properties, by comparing co-temporal data taken by the various instruments,} such that the apparent surface coverage by faculae in the periods of overlap is similar. The model utilizes the intensity spectra of quiet Sun, faculae and sunspot umbra and penumbra from \cite{unruh99}, which is actually a significant update of the \cite{solanki98} intensity contrasts employed in NRLSSI. {The disc-integrated solar spectrum at the time of each magnetogram and intensity image pair is given by the sum of the \cite{unruh99} intensity spectra, weighted by the apparent solar surface coverage of faculae and sunspots in the data pair.}

{There is one free parameter in the model. The magnetic concentrations making up network and faculae are largely unresolved in available full-disc magnetograms. As such, the proportion of each resolution element occupied by such features (the so-termed `filing factor') has to be inferred indirectly from the magnetogram signal. By modulating the empirical relationship between the filling factor and the magnetogram signal \citep{fligge00}, the free parameter scales the amplitude of the facular brightening contribution to solar irradiance variability. The appropriate value of the free parameter was determined by comparing reconstructed TSI to observed TSI.}

The radiative transfer code used to derive the intensity spectra, ATLAS9, assumes local thermodynamic equilibrium (LTE), which is not realistic at the height in the solar atmosphere where prominent spectral lines, in particular in the UV, are formed. {This is compounded by the use of opacity distribution functions (ODFs) to represent spectral lines in the code \citep[see discussion in][]{yeo14c}.} As a result, the absolute level in reconstructed spectra is markedly lower than in observations below 300 nm \citep{krivova06,yeo14a}. To account for this, the 180 to 300 nm segment of the reconstruction is offset to the WHI reference solar spectra \citep{woods09}, thus taking the absolute level of the WHI spectra whilst retaining the variability from the model. {The offset was determined by comparing the WHI reference spectra, of which there are three, to reconstructed SSI over the periods the WHI spectra were taken.} The 115 to 180 nm segment is offset and linearly scaled to match the SORCE/SOLSTICE record. It is worth noting that in doing so, only the absolute scale of the SORCE/SOLSTICE record is adopted, the time variation from the model is preserved.

{The present reconstruction procedure, summarized above, differs from the preceding version of SATIRE-S \citep{ball14b} in two key aspects. One, how the model output based on observations from different instruments are combined together and two, the correction in the UV for the LTE and ODF approximations. Instead of cross-calibrating the various magnetogram data sets, \cite{ball14b} had adapted the parameters of the model to each data set. This left weak but palpable discrepancies in the reconstructed solar irradiance based on data from different instruments, which was then corrected for by regression. In cross-calibrating the magnetogram data sets, the model output from the various data sets are now mutually consistent without the need for any such correction. \cite{ball14b} corrected their reconstruction for the LTE and ODF approximations by offsetting and linearly scaling the 115 nm to 270 nm segment to match the UARS/SUSIM record. \cite{yeo14b} matched the 115 to 180 nm segment to SORCE/SOLSTICE SSI in a similar manner but only offset the 180 to 300 nm segment to the WHI reference spectra on the observation that at these wavelengths, an offset is sufficient. The goal was to correct for the LTE and ODF approximations while introducing as few empirical corrections as practically possible. The results of the two approaches are not significantly different. As we will see in Sect. \ref{review}, the SATIRE-S reconstruction, now independent of the UARS/SUSIM record, is a very good match to this particular record.}

{The intensity spectra of solar surface features employed in SATIRE-S, as with all other present-day semi-empirical models, are based on one-dimensional models of the solar atmosphere, which do not necessarily represent the true laterally-averaged property of the inhomogeneous solar atmosphere \citep{uitenbroek11,holzreuter13}. This, along with the use of an empirical relationship to infer the faculae filling factor from the magnetogram signal and the LTE and ODF simplifications in ATLAS9 represent the key limitations of the model. They are discussed in greater detail, along with the proposed future work to address them, in \cite{yeo14a,yeo14c}.}

\subsection{Morrill et al. 2011 (Mea11)}
\label{modelsmea11}

One of the more striking distinctions between the NRLSSI and SATIRE-S models is the weaker solar cycle variability in the UV produced by the former. {It has been suggested that in confining to the rotational variability in the regression of the PSI and the Mg II index to the UARS/SOLSTICE record, solar cycle variability might be underestimated in NRLSSI \citep{lean00,ermolli13,yeo14a}.} To investigate this {claim}, we will also examine the empirical reconstruction of UV SSI by Mea11. The reconstruction, extending from 1 to 410 nm, is based on relating the Mg II index to the SSI records from TIMED/SEE \citep[1 to 150 nm segment of the model,][]{woods05} and UARS/SUSIM \citep[150 to 410 nm,][]{brueckner93,floyd03}. In contrast to the approach taken by NRLSSI, the Mg II index is regressed to measured SSI directly, that is, without first detrending the index and SSI time series. As we will see in the following section, the {UARS/SUSIM} record, unlike the UARS/SOLSTICE record, is stable enough to reveal solar cycle variability even at wavelengths longwards of 220 nm, allowing a direct regression of the Mg II index.

\section{Wavelength Dependence of Solar Irradiance Variability}
\label{ssi}

\subsection{Critical Review of Measurements and Models}
\label{review}

\begin{figure}
\noindent\includegraphics[width=\textwidth]{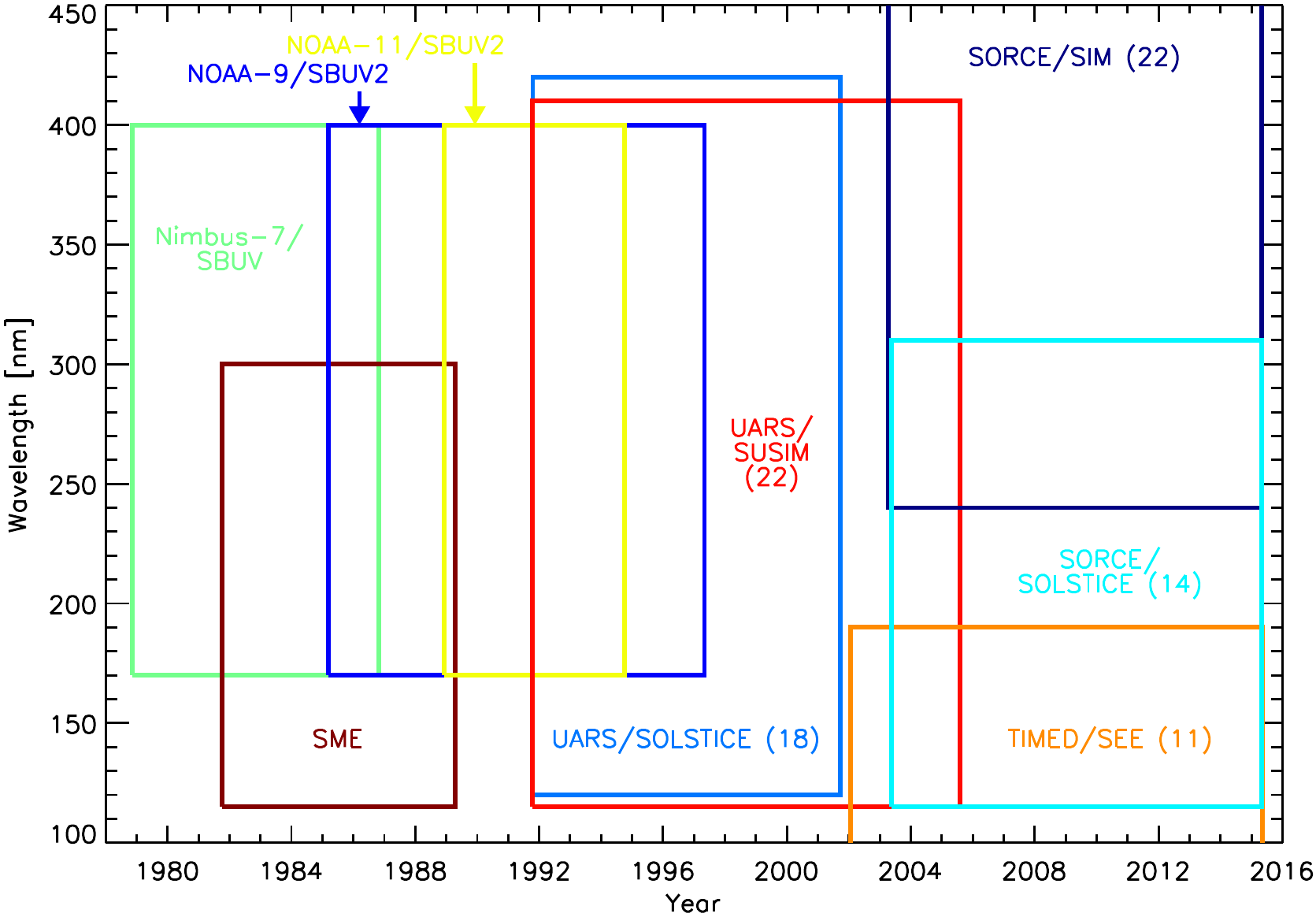}
\caption{The time and wavelength coverage of the extended{, daily-cadence} SSI records calibrated for instrument degradation (latest version number at the time of writing in parenthesis). The SSI integrated over 170 to 180 nm and 290 to 300 nm is illustrated in Fig. \ref{absolute}.}
\label{coverage}
\end{figure}

\begin{figure}
\noindent\includegraphics[width=\textwidth,natwidth=2023,natheight=2266]{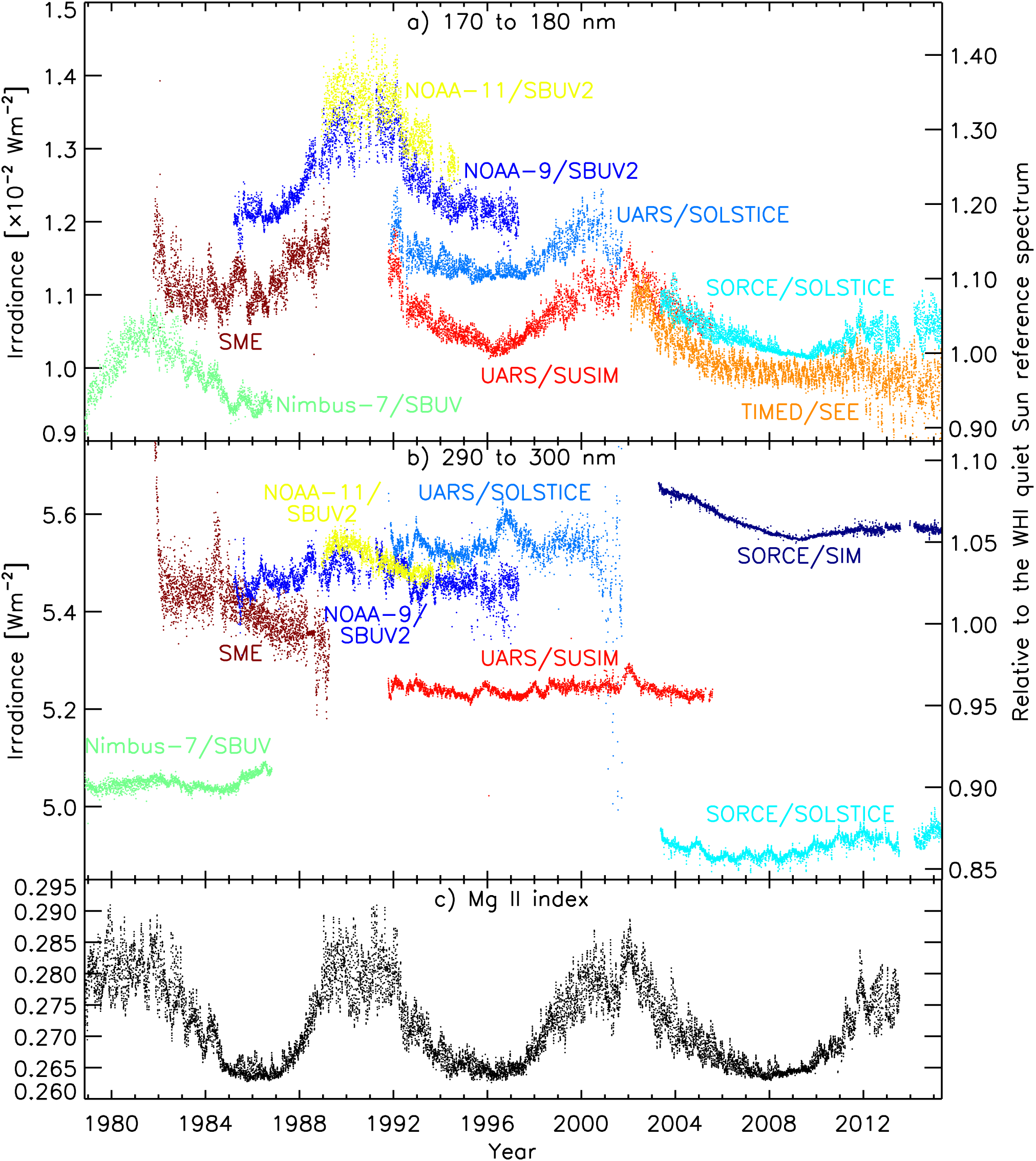}
\caption{Integrated SSI over a) 170 to 180 nm and b) 290 to 300 nm in units of ${\rm Wm^{-2}}$ on the left axis and relative to the WHI quiet Sun reference spectrum \citep{woods09} on the right axis. c) The LASP Mg II index composite \citep{snow05b}.}
\label{absolute}
\end{figure}

{As noted in the introduction, SSI between 120 and 400 nm has been monitored since 1978 (Figs. \ref{coverage} and \ref{absolute}). In this study, we examine the extended (at least a few years), daily-cadence satellite records that are calibrated for instrument degradation, listed in Table \ref{ssirecords}. The period covered by each record and the spectral range and resolution are also listed.}

\begin{figure}
\noindent\includegraphics[width=.75\textwidth]{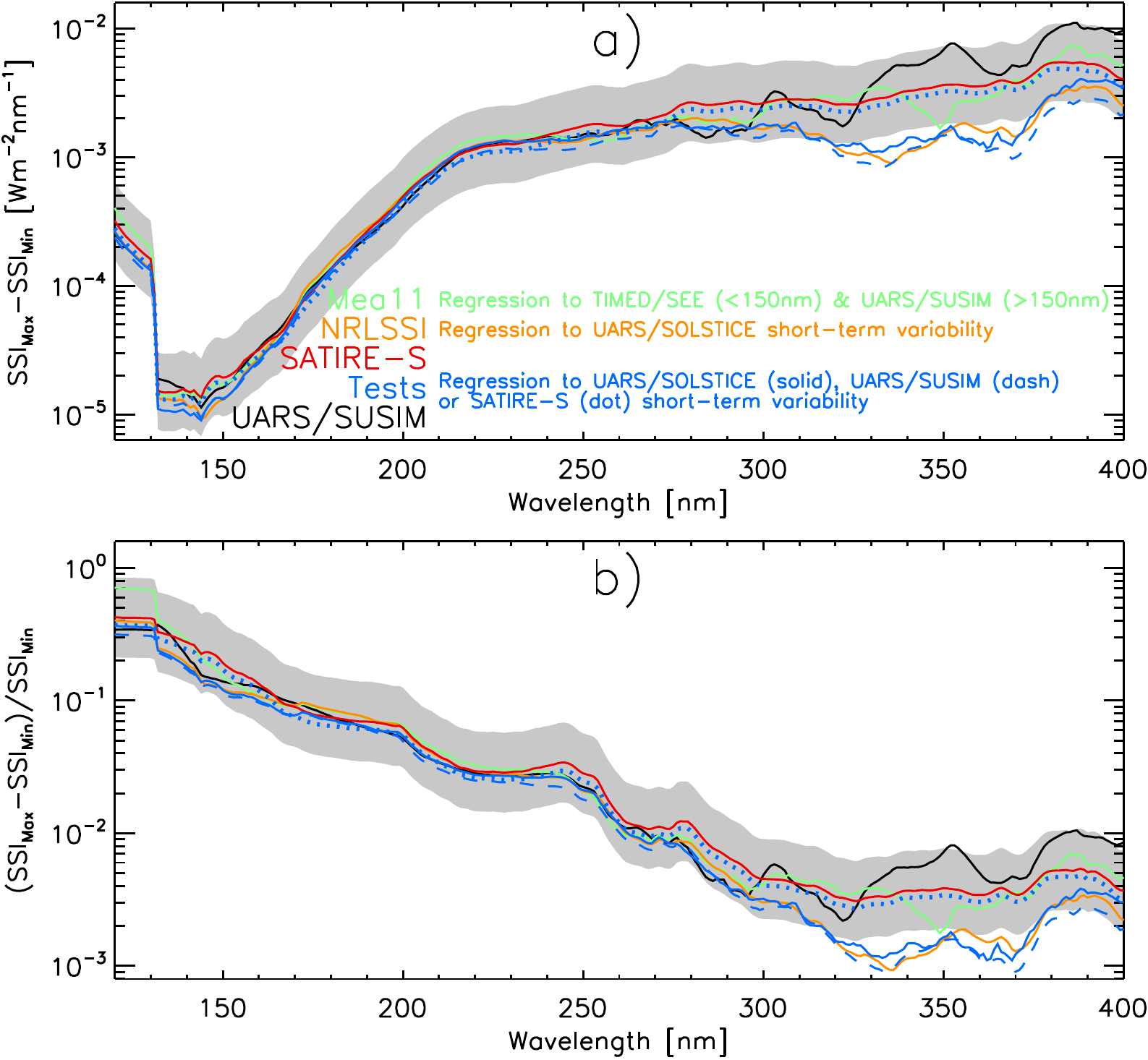}
\caption{The change in SSI between the 1996 solar cycle minimum and the 2000 maximum, as a function of wavelength, in the SUSIM record (black), the Mea11 (green), NRLSSI (orange) and SATIRE-S (red) reconstructions, as well as the test reconstructions based on the regression of the PSI and Mg II index to the rotational variability in SSI (blue, see Sect. \ref{testreconstructions}). The variation in a) absolute terms and b) as a proportion of the 1996 minimum level is depicted. We considered the mean SSI over a 20 nm window centred on each wavelength. The elevated levels between 120 and 130 nm arise from the large solar cycle variability at the Lyman-$\alpha$ line (to the order of $10^{-3}\:{\rm Wm^{-2}nm^{-2}}$ or a few tenths of the overall level). To elucidate the scale of the scatter between the various data sets, we shaded the range bounded by $50\%$ and $200\%$ of the level in the SATIRE-S reconstruction. Here, and in the rest of the study, we adopt the epoch of solar cycle minima and maxima defined by NOAA (www.ngdc.noaa.gov/stp/space-weather/solar-data/solar-indices/sunspot-numbers/cycle-data/).}
\label{solarcycleamplitude}
\end{figure}

Between 120 and 400 nm, the absolute variation in SSI over the solar cycle increases by over two orders of magnitude with wavelength (Fig. \ref{solarcycleamplitude}a). However, in relative terms, this variability actually diminishes by at least two orders of magnitude across this spectral region (Fig. \ref{solarcycleamplitude}b). This is at least partly why the uncertainty in the variation over the solar cycle and instrumental artefacts such as spurious spikes/dips become increasingly apparent in the body of UV SSI observations towards longer wavelengths (compare Figs. \ref{absolute}a and \ref{absolute}b).

\begin{figure}
\noindent\includegraphics[width=.8\textwidth]{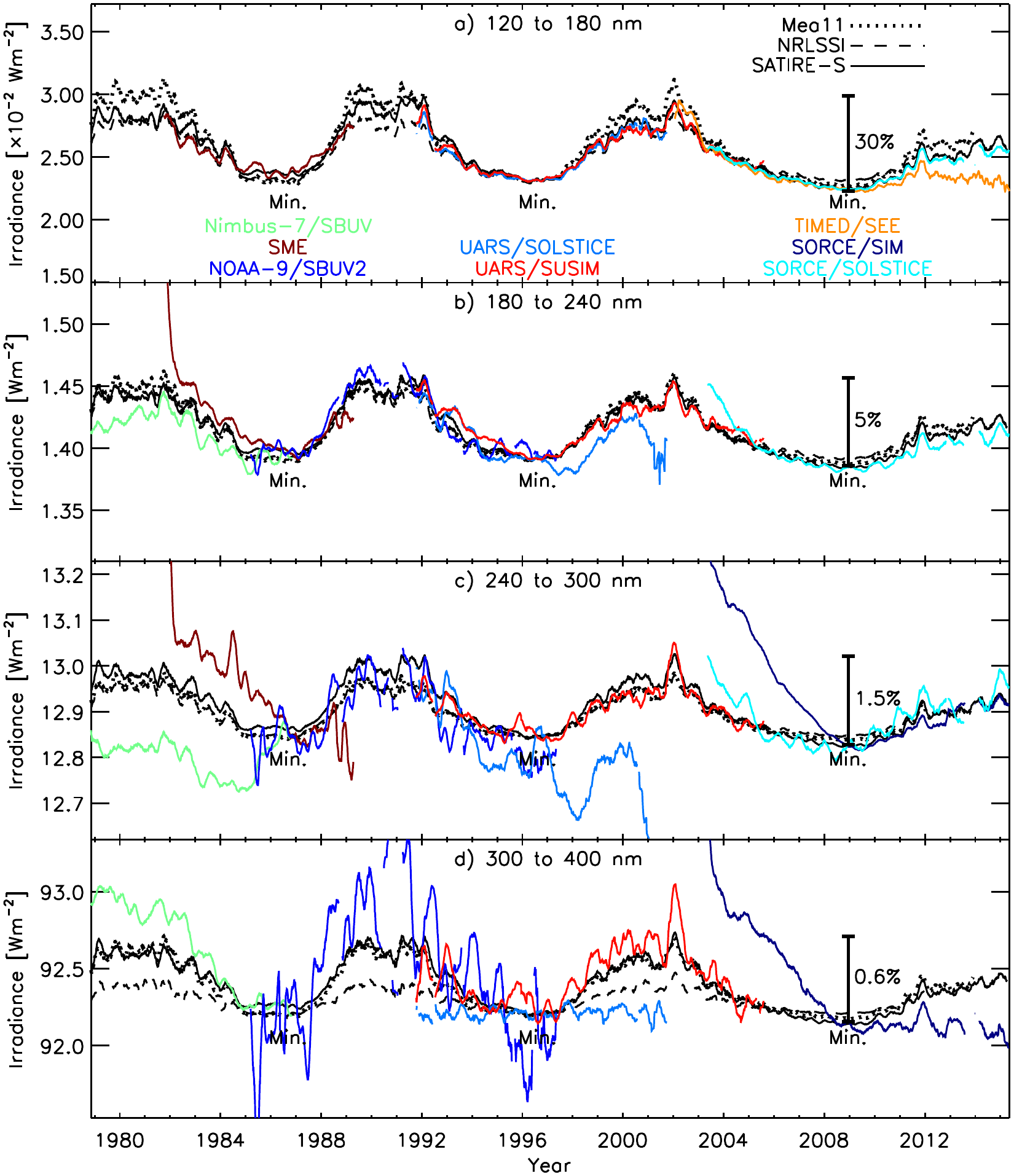}
\caption{The integrated SSI within the annotated wavelength intervals in measurements (colour) and produced by models (black, see legend in top panel), against time. The SSI time series were smoothed by taking the 81-day moving average. Solar cycle minima are indicated. The various time series are normalized to the SATIRE-S reconstruction (black solid) at solar cycle minimum (1986 for Nimbus-7/SBUV, SME and NOAA-9/SBUV2, 1996 for the UARS records and the Mea11 and NRLSSI reconstructions, and 2008 for the {TIMED/SEE} and the SORCE records). The gaps correspond to periods longer than 27 days, about a solar rotation period, with no data.}
\label{uvssi}
\end{figure}

In Fig. \ref{uvssi}, we bring the SSI records that encompass at least one solar cycle minimum to a common scale by normalizing them to the SATIRE-S reconstruction at the minimum. To simplify the discussion, we focus on the integrated SSI over 120 to 180 nm, 180 to 240 nm, 240 to 300 nm and 300 to 400 nm. These intervals were selected on the observation that the overall agreement between the various records starts to deteriorate going above around 180 nm, discussed below, and taking into account that the {SORCE/SIM} record only extends down to 240 nm while the SME and SORCE/SOLSTICE records extend up to around 300 nm (Fig. \ref{coverage}).

In the 120 to 180 nm interval, the various records exhibit solar cycle modulation and are broadly consistent with one another where they overlap (Fig. \ref{uvssi}a){, with the exception of TIMED/SEE, discussed later in this section}. Between 180 and 240 nm, available records still show solar cycle modulation but at varying amplitudes such that they start to differ significantly in the periods they overlap (Fig. \ref{uvssi}b). Going above 240 nm, certain records no longer vary with the solar cycle in a consistent manner (Figs. \ref{uvssi}c and \ref{uvssi}d).

It is established that TSI is correlated to the 11-year solar magnetic cycle. In order for SSI to exhibit non-cyclic trends over the solar cycle, the underlying mechanism must be such that the integral of this component of the variability over all wavelengths is too weak to be noticeable in TSI. There is no obvious candidate mechanism that can drive the radiant output of the Sun in such a manner. Consequently, it is doubtful that the non-cyclic trends over the solar cycle noted in certain SSI records {longwards of 240 nm} are solar in origin.

The deteriorating agreement between the various records and the emergence of non-cyclic trends over the solar cycle towards longer wavelengths is likely a consequence of the weakening of the underlying solar cycle variability to below the limits of measurement stability.

\begin{figure}
\noindent\includegraphics[width=\textwidth]{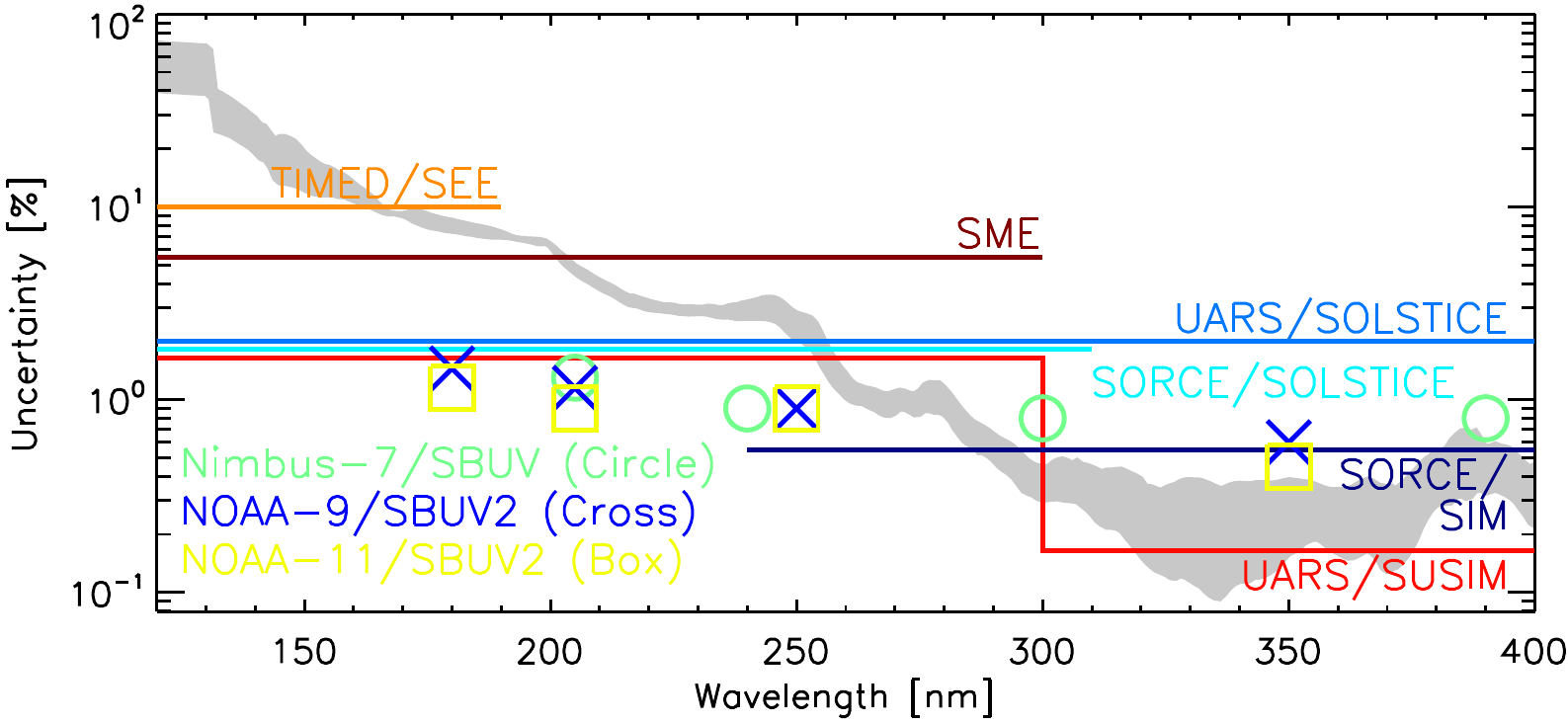}
\caption{The 1-$\sigma$ uncertainty in the variation over the solar cycle in SSI records as a percentage of the overall level. For comparison, we shaded the range of values for the change in SSI over the ascending phase of solar cycle 23 in the Mea11, NRLSSI and SATIRE-S reconstructions (from Fig. \ref{solarcycleamplitude}b). {The uncertainty is given by the long-term uncertainty of the respective instruments reported in the literature or in the data header (see Table \ref{reportedlongtermuncertainty} and text)}. We represent the values for the three SBUV instruments using different plot symbols and offset the SORCE/SOLSTICE line, which would otherwise overlap with the UARS/SOLSTICE line, down slightly by $0.1\%$ to aid visibility.}
\label{longtermuncertainty}
\end{figure}

{In Fig. \ref{longtermuncertainty}, we show the uncertainty in the variation over the solar cycle of each SSI record, given by the reported long-term uncertainty (listed in Table \ref{reportedlongtermuncertainty}). An uncertainty range is reported for SORCE/SOLSTICE (0.2 to 0.5$\%$/year). In this case, we considered the mid-point of the cited range (0.35$\%$/year). For the records where the long-term uncertainty is given in units of percent per year, we multiplied this per annum uncertainty by 5.5, around half the typical length of a solar cycle, to arrive at the estimate of the uncertainty in the variation over the solar cycle depicted in Fig. \ref{longtermuncertainty}.}

This quantity is a crude indication of the weakest solar cycle modulation detectable. For example, over the ascending phase of solar cycle 23, the variation in UV SSI (as indicated by the Mea11, NRLSSI and SATIRE-S reconstructions, shaded grey in {Fig. \ref{longtermuncertainty}}) declines to below the uncertainty of the Nimbus-7/SBUV, SME, NOAA-9/SBUV2, UARS/SOLSTICE and {SORCE/SIM} records between 200 and 300 nm. This is roughly where we found the overall trend over the solar cycle in these records to begin to turn non-cyclic (Fig. \ref{uvssi}). {For the {UARS/SUSIM} record, the uncertainty is smaller or close to the amplitude of the solar cycle at most wavelengths and this record exhibits solar cycle modulation consistently over its entire wavelength range.}

{The only exceptions to this broad pattern in the body of SSI observations are the SORCE/SOLSTICE and TIMED/SEE records. For SORCE/SOLSTICE,} although solar cycle variability weakens to below the uncertainty above around 250 nm (Fig. \ref{longtermuncertainty}), the record continues to show solar cycle modulation (Fig. \ref{uvssi}c). This implies that the long-term uncertainty of this record might be, at least above 250 nm, smaller than reported. {For TIMED/SEE, even though the uncertainty is below or at least comparable to solar cycle variability (Fig. 5), the record exhibits little overall variability between the 2008 minimum and 2015, rising only weakly before declining back to 2008 levels (Fig. 4a). The more recent TIMED/SEE measurements might be affected by unresolved instrumental trends.}

The overall trend in the Nimbus-7/SBUV, SME, NOAA-9/SBUV2 and UARS/SOLSTICE records is relatively well-replicated by Mea11, NRLSSI and SATIRE-S up to the wavelengths where measurement uncertainty starts to dominate (compare Figs. \ref{uvssi} and \ref{longtermuncertainty}). {While earlier versions of the SORCE/SOLSTICE record exhibit a stronger decline between 2003 and 2008 than replicated in models (see Sect. \ref{introduction}), this is no longer the case with the latest revision (version 14, dated May 2015).} For {UARS/SUSIM} and SORCE/SOLSTICE, the records stable enough to reveal solar cycle modulation consistently over their respective wavelength range, the overall trend is largely reproduced, with a notable exception. Between 300 and 400 nm, the amplitude of the variation over the solar cycle in the NRLSSI reconstruction is only about half that indicated by {UARS/SUSIM}, which is reasonably well replicated in the two other reconstructions (Figs. \ref{solarcycleamplitude} and \ref{uvssi}d). {In this wavelength interval, there is a quasi-annual oscillation in the UARS/SUSIM record that is most likely instrumental in origin. It is difficult to establish, to better than the amplitude of this quasi-annual wobble, the goodness of the agreement between this record and the SATIRE-S reconstruction. The alignment between the SATIRE-S and Mea11 reconstructions here does, however, confer confidence to the solar cycle variability replicated by SATIRE-S. Recall, above 150 nm, the Mea11 reconstruction is essentially a correction of the {UARS/SUSIM} record to the stability of the Mg II index \citep{marchenko14}.}

Apart from the weaker solar cycle variability between 300 and 400 nm produced by NRLSSI, the reconstructions examined are broadly consistent with one another (Figs. \ref{solarcycleamplitude} and \ref{uvssi}). The solar cycle variability in NRLSSI is already weaker than in SATIRE-S between 240 and 300 nm. This difference is, however, too small to discriminate between the two reconstructions with the UARS/SUSIM and SORCE/SOLSTICE records, the two records stable enough to exhibit solar cycle modulation in this wavelength interval.

As noted in earlier studies for previous versions of the {SORCE/SIM} record (see Sect. \ref{introduction}), the latest revision {(version 22, dated May 2015)} indicates a decline in 240 to 400 nm SSI between 2003 and 2008 that is grossly steeper than what is recorded by SORCE/SOLSTICE or replicated by the models examined (Fig. \ref{uvssi}). Numerous studies have concluded that this acute drop registered by {SORCE/SIM} might not be real due to unresolved instrumental trends \citep{ball11,deland12,lean12,unruh12,ermolli13,morrill14,yeo14a,yeo14b}. This assertion is supported here by the agreement between the concurrent measurements from SORCE/SOLSTICE and the various reconstructions (Fig. \ref{uvssi}c).

In this discussion, we have not considered the fact that, due to the varying approaches and observations accounted for in the derivation, the absolute level in the Mea11, NRLSSI and SATIRE-S reconstructions are offset from one another. Between 120 and 400 nm, the offset between the various reconstructions is, except at the lowest wavelengths, largely within $10\%$. This spread in the absolute level is weak such that there is little qualitative difference in how the solar cycle variability in these reconstructions compare to one another and to SSI records whether we look at it in absolute (Fig. \ref{solarcycleamplitude}a) or relative terms (Fig. \ref{solarcycleamplitude}b).

The SATIRE-S reconstruction is consistent with UV SSI records where the latter is stable enough to reveal solar cycle modulation and also with the independent reconstruction from Mea11. This renders strong support to the solar cycle variability in the UV produced by the SATIRE-S model. {It is worth emphasizing here that while the 115 to 180 nm segment of the SATIRE-S reconstruction is offset and scaled to match SORCE/SOLSTICE SSI, between 180 to 400 nm, the solar irradiance variability returned by the model is not altered in any manner. The 180 to 300 nm segment is only offset to the WHI reference spectra. The empirical corrections for the LTE and ODF approximations has no influence on the agreement in the solar cycle variability between SATIRE-S and the various UV SSI records and the Mea11 reconstruction longwards of 180 nm.}

\begin{figure}
\noindent\includegraphics[width=\textwidth]{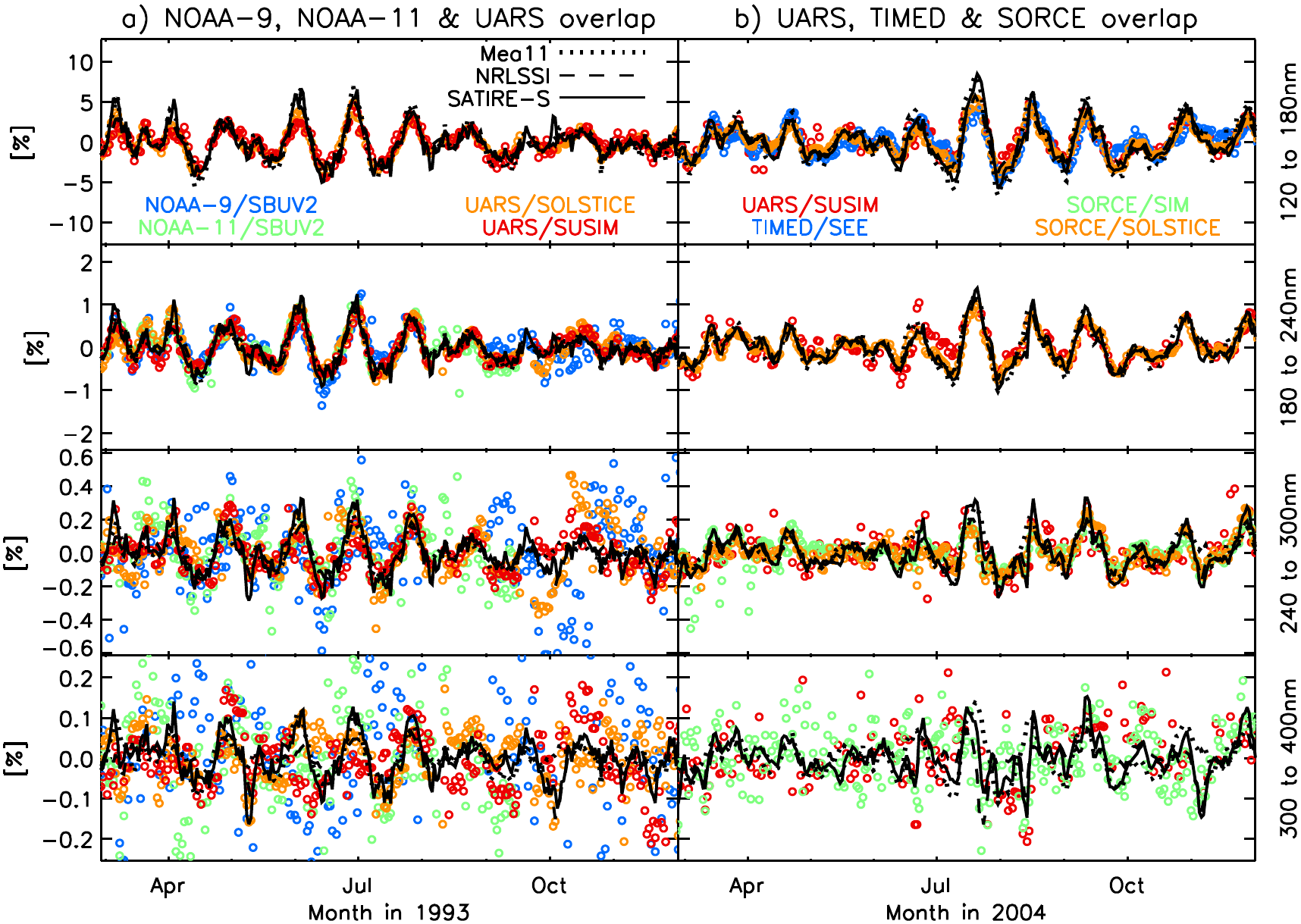}
\caption{The rotational variability in the integrated SSI in the annotated wavelength intervals (right-hand-side) over a 9-month period in the overlap between a) the NOAA-9, NOAA-11 and UARS missions, and between b) the UARS, TIMED and SORCE missions. The coloured circles correspond to values from satellite records and the black lines to the Mea11 (dot), NRLSSI (dash) and SATIRE-S (solid) reconstructions, largely overlapping at shorter wavelengths due to the close similarity. Rotational variability, expressed in percentage deviation from the overall level, is isolated by dividing each integrated SSI time series by the corresponding 81-day moving average.}
\label{uvssirotationalvsmodels}
\end{figure}

As with the variation over the solar cycle, rotational variability, in relative terms, weakens with wavelength such that measurement uncertainty becomes increasingly dominant. This is evident in the growing discrepancy between the concurrent measurements from the various instruments towards longer wavelengths (Fig. \ref{uvssirotationalvsmodels}). The models examined replicate the well-observed rotational variability in the 120 to 180 nm and 180 to 240 nm intervals, where available records are consistent with one another. Longwards of 240 nm, as the consistency between the various records deteriorates, so does the agreement between measurements and models. The worsening reliability of measured rotational variability towards longer wavelengths has implications for models which rely on it, such as the NRLSSI.

\subsection{The Discrepant Solar Cycle Variability in the NRLSSI and SATIRE-S Reconstructions}
\label{testreconstructions}

Compared to SATIRE-S, NRLSSI produces significantly weaker solar cycle variability in the UV longwards of around 300 nm. As stated in Sect. \ref{modelsmea11}, this {has been} suggested to arise from NRLSSI confining the regression of the PSI and the Mg II index to the UARS/SOLSTICE record to the short-term variability by first detrending each time series. As the long-term uncertainty of the UARS/SOLSTICE record is severe enough to obscure solar cycle modulation at longer wavelengths, it is not possible to examine the effects of this step taken by NRLSSI by looking at the result of fitting the index data to the record directly (i.e., without first detrending each time series). In this study, we make use of the fact that the {UARS/SUSIM} record, which extends a similar period and wavelength range as the UARS/SOLSTICE record, is stable enough to reveal clear solar cycle modulation over the entire wavelength range of the instrument (Fig. \ref{uvssi}). This allows us to probe the NRLSSI approach by applying it to the {UARS/SUSIM} record and comparing the resultant reconstruction to the {UARS/SUSIM} record and the Mea11 reconstruction.

To this end, we generated three test reconstructions, based on the regression of the PSI composite by \cite{balmaceda09} and the LASP Mg II index composite \citep{snow05b} to each of the two UARS records and the SATIRE-S reconstruction. We restricted the regression to the short-term variability by subtracting, from each time series, the corresponding 81-day moving average prior to fitting. Each test reconstruction is then completed by applying the regression coefficients to the original index data and offsetting the result to the absolute level in the SSI time series referenced at the 1996 solar cycle minimum.

\begin{figure}
\noindent\includegraphics[width=.95\textwidth]{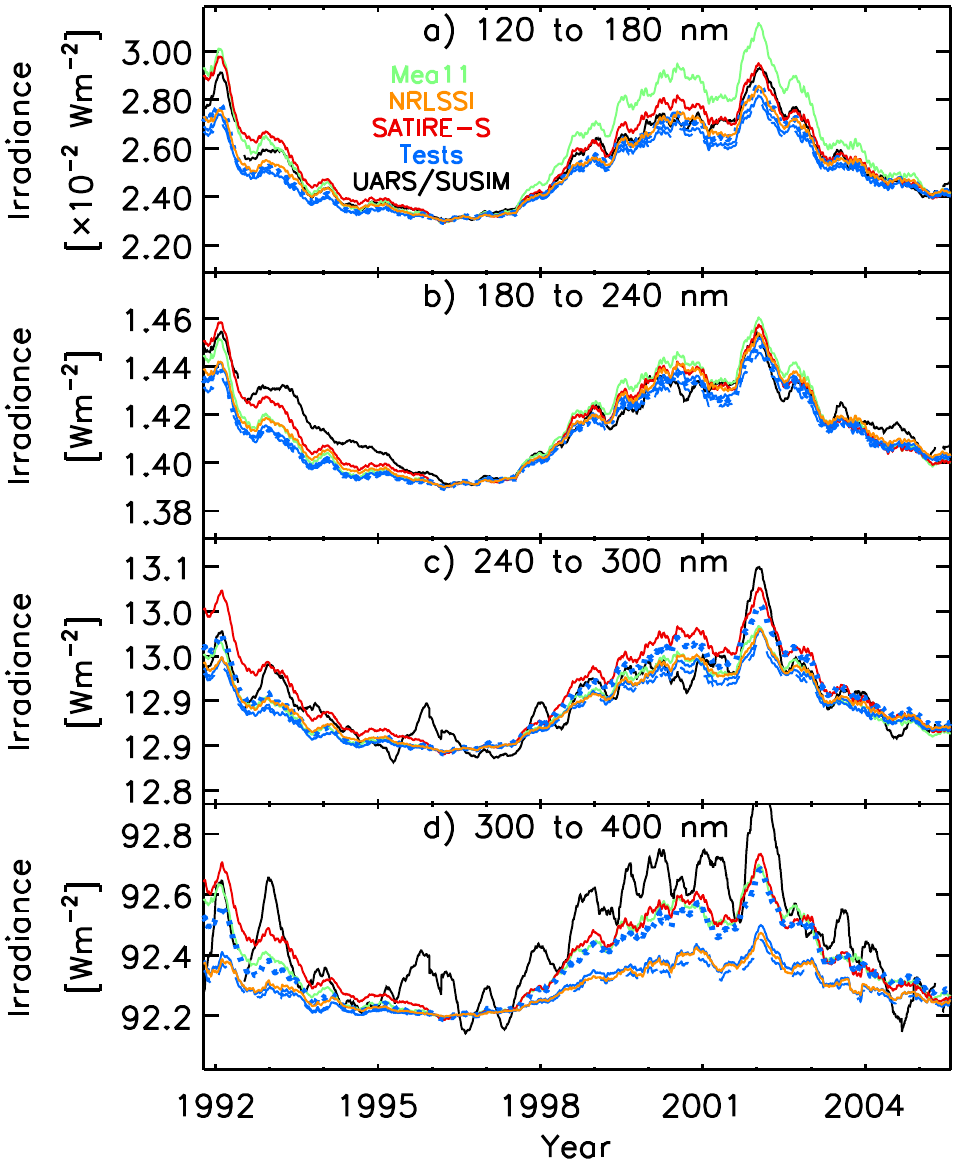}
\caption{Similar to Fig. \ref{uvssi}, except restricted to the duration of the SUSIM record. The blue time series correspond to the test reconstructions based on the regression of the PSI and the Mg II index to the rotational variability in UARS/SOLSTICE (solid), SUSIM (dashed) and SATIRE-S SSI (dotted).}
\label{comparemodels}
\end{figure}

\begin{figure}
\noindent\includegraphics[width=\textwidth]{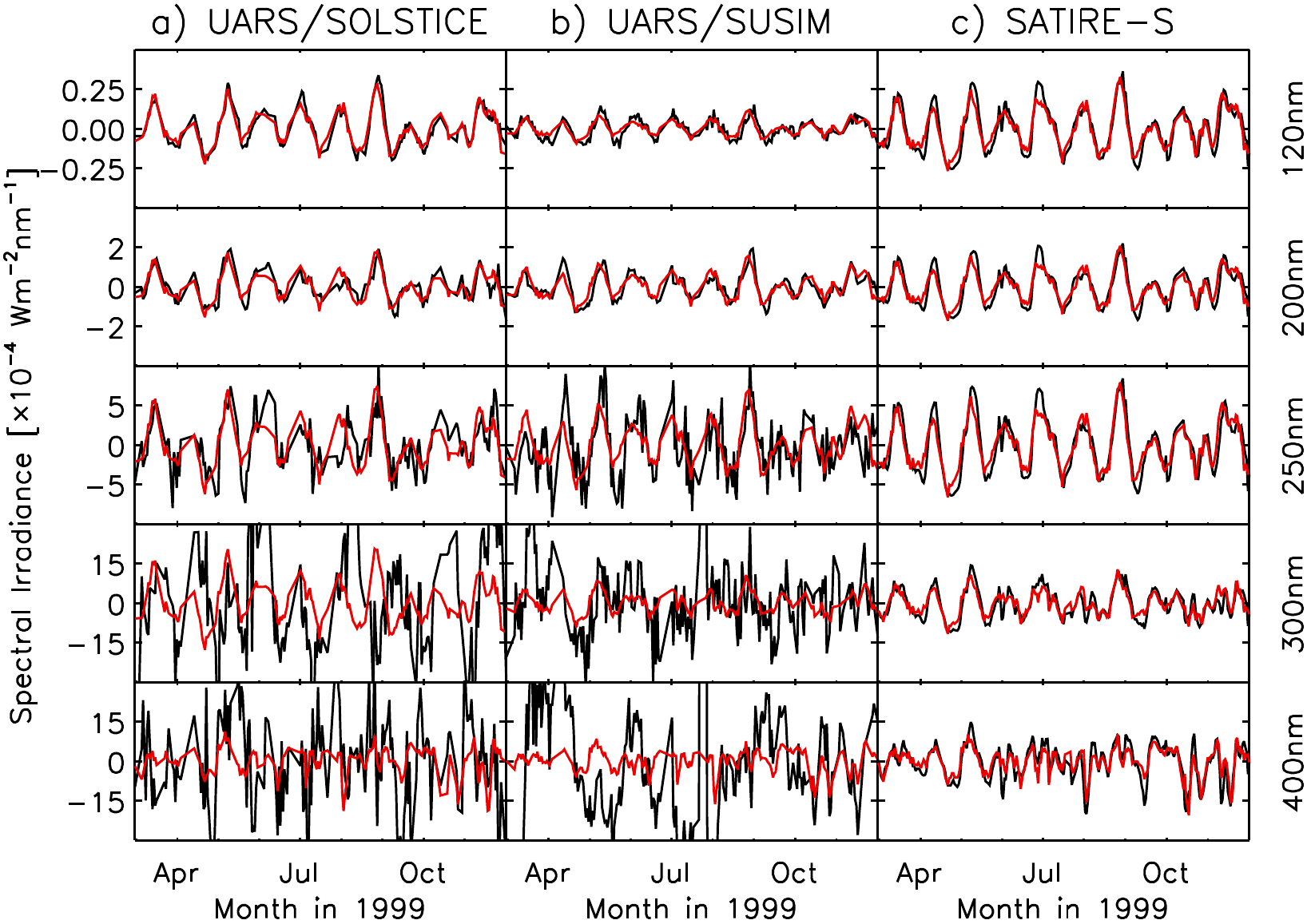}
\caption{Black: Detrended SSI from the a) UARS/SOLSTICE and b) SUSIM records and the c) SATIRE-S reconstruction at the wavelengths indicated (right-hand-side) over a 9-month period in 1999. Red: The multiple linear regression of the detrended PSI and Mg II index.}
\label{testmodels}
\end{figure}

The wavelength dependence of the variation over the solar cycle in the test reconstruction referenced to the UARS/SOLSTICE record (blue solid lines, Figs. \ref{solarcycleamplitude} and \ref{comparemodels}) is nearly identical to the NRLSSI reconstruction (green). This verifies that the procedure taken to generate the test reconstructions is largely analogous to the NRLSSI. The solar cycle variability in the test reconstruction referenced to the {UARS/SUSIM} record (blue dashed lines) is similarly compatible to the NRLSSI reconstruction. Applying the NRLSSI approach to the {UARS/SUSIM} record results in the reconstructed solar cycle variability being weaker, towards longer wavelengths, than in the {UARS/SUSIM} record (black) and the Mea11 reconstruction (yellow). Possible causes of this discrepancy is alluded to by the following observations.
\begin{enumerate}
	\item The agreement between the detrended UARS records (black, Figs. \ref{testmodels}a and \ref{testmodels}b) and the regression of the detrended PSI and Mg II index (red) deteriorates with wavelength. The agreement is especially poor longwards of 300 nm, which is incidentally the wavelength range where the solar cycle variability in the test reconstructions built on the rotational variability in these records, like NRLSSI, diminishes to about half the amplitude indicated by the {UARS/SUSIM} record and the Mea11 and SATIRE-S reconstructions (Figs. \ref{solarcycleamplitude} and \ref{comparemodels}). Clearly, this is related to the deteriorating reliability of the rotational variability in UV SSI measurements towards longer wavelengths, discussed in the previous section.
	\item The regression of the detrended PSI and Mg II index to the detrended SATIRE-S reconstruction is a good fit (Fig. \ref{testmodels}c) and the solar cycle variability in the resultant reconstruction  (blue dotted lines, Figs. \ref{solarcycleamplitude} and \ref{comparemodels}) is consistent with the SATIRE-S reconstruction (red) even above 300 nm.
\end{enumerate}
Together, these observations suggest that there is nothing wrong with the NRLSSI approach in and of itself but longwards of 300 nm, the rotational variability in the UARS records is no longer accurate enough to establish the relationship between indices of magnetic activity and SSI reliably. Consequently, the solar cycle variability produced by NRLSSI is too weak between 300 and 400 nm.

It is worth emphasizing here the distinction between the short-term sensitivity and long-term stability of SSI records, the uncertainty of which arise from different underlying physical processes \citep[see, for example,][]{cebula98}. The observation here that the short-term variability in the {UARS/SUSIM} record is too noisy to reconstruct SSI by the NRLSSI approach does not conflict with the fact that the record is stable enough to reveal solar cycle modulation throughout its wavelength range.

We generated similar test reconstructions taking the Nimbus-7/SBUV, NOAA-9/SBUV2 and {SORCE/SIM} records as the reference, not shown here to avoid repetition. As with the test reconstructions based on the UARS records, in each instance, the agreement between measured rotational variability and the regression of the index data deteriorates with wavelength and the reconstructed solar cycle variability is much closer to the level in NRLSSI than to that in the {UARS/SUSIM} record and the Mea11 and SATIRE-S reconstructions. 

\section{Summary and Conclusions}
\label{conclusions}

TSI and UV SSI have been measured from space, almost without interruption, since 1978 through a succession of satellite missions. At the same time, models aimed at reconstructing solar irradiance by relating the variability to solar magnetic activity were developed. The NRLSSI and SATIRE-S models provide the most comprehensive, in terms of time and wavelength coverage, reconstructions of TSI and SSI over the period of satellite observation currently available.

Although considerable progress has been made in the measurement and modelling of solar irradiance variability, there remains controversy over the wavelength dependence of the variation over the solar cycle in the UV, with repercussions on our understanding of the influence of solar irradiance variability on the stratosphere. In this paper, we presented a critical review of solar irradiance variability in the UV since 1978. The emphasis is on the wavelength dependence of solar cycle variability in observations and in the NRLSSI and SATIRE-S reconstructions. Our aim was to validate the variability produced by the two models and therefore their utility for climate studies.

In the UV, solar irradiance variability, in relative terms, diminishes with wavelength. This has the effect that observed variability is increasingly dominated by measurement uncertainty towards longer wavelengths. The solar cycle variability replicated by SATIRE-S is consistent with UV SSI records at the wavelengths where it is not obscured in the latter by measurement uncertainty. It is also supported by the independent, empirical reconstruction of UV SSI by \cite{morrill11}. The weaker solar cycle variability produced by NRLSSI between 300 and 400 nm is not evident in any available record. This was shown to have likely arisen from the short-term uncertainty in the SSI record referenced in the reconstruction, that from UARS/SOLSTICE. Solar cycle variability is similarly underestimated in test reconstructions derived by a procedure analogous to NRLSSI using other SSI records as the reference.

The NRLSSI approach is sound in principle but reconstructed solar cycle variability in the UV is detrimentally affected by the short-term uncertainty in the SSI observations employed in the derivation of the reconstruction. The solar cycle variability in the UV produced by SATIRE-S is well-supported by observations. {Between the two models, we recommend the reconstruction from SATIRE-S for climate studies.}

\begin{acknowledgements}
We are thankful to G\'{e}rard Thuillier for helpful discussions. Appreciation goes to the individuals and teams responsible for the measurements and models featured in this work; Mathew Deland (SBUV SSI), Judith Lean (NRLSSI model), Marty Snow (LASP Mg II index composite) and Tom Woods (WHI reference spectra). We also made use of SSI observations from the SME, UARS, SORCE and TIMED missions. This work is partly supported by the German Federal Ministry of Education and Research under project 01LG1209A, a STSM grant from COST Action ES1005 `TOSCA' and the Ministry of Education of Korea through the BK21 plus program of the National Research Foundation. {The \cite{morrill11} reconstruction is available on request (jeff.morrill@nrl.navy.mil). The SBUV records are available at sbuv2.gsfc.nasa.gov/solar/, the UARS/SUSIM record at wwwsolar.nrl.navy.mil/uars/ and the \cite{balmaceda09} PSI composite and the SATIRE-S reconstruction at www2.mps.mpg.de/projects/sun-climate/data.html. The LISIRD data centre (lasp.colorado.edu/lisird/) hosts the LASP Mg II index composite, the NRLSSI reconstruction, the WHI reference spectra and the SME, UARS/SOLSTICE, SORCE and TIMED/SEE records.}
\end{acknowledgements}


\begin{thebibliography}{}
\providecommand{\natexlab}[1]{#1}
\expandafter\ifx\csname urlstyle\endcsname\relax
  \providecommand{\doi}[1]{doi:\discretionary{}{}{}#1}\else
  \providecommand{\doi}{doi:\discretionary{}{}{}\begingroup
  \urlstyle{rm}\Url}\fi

\bibitem[{\textit{{Ball} et~al.}(2011)\textit{{Ball}, {Unruh}, {Krivova},
  {Solanki}, and {Harder}}}]{ball11}
{Ball}, W.~T., Y.~C. {Unruh}, N.~A. {Krivova}, S.~K. {Solanki}, and J.~W.
  {Harder} (2011), {Solar irradiance variability: a six-year comparison between
  SORCE observations and the SATIRE model}, \textit{Astron. Astrophys.},
  \textit{530}, A71, \doi{10.1051/0004-6361/201016189}.

\bibitem[{\textit{{Ball} et~al.}(2012)\textit{{Ball}, {Unruh}, {Krivova},
  {Solanki}, {Wenzler}, {Mortlock}, and {Jaffe}}}]{ball12}
{Ball}, W.~T., Y.~C. {Unruh}, N.~A. {Krivova}, S.~K. {Solanki}, T.~{Wenzler},
  D.~J. {Mortlock}, and A.~H. {Jaffe} (2012), {Reconstruction of total solar
  irradiance 1974-2009}, \textit{Astron. Astrophys.}, \textit{541}, A27,
  \doi{10.1051/0004-6361/201118702}.

\bibitem[{\textit{{Ball} et~al.}(2014{\natexlab{a}})\textit{{Ball}, {Mortlock},
  {Egerton}, and {Haigh}}}]{ball14a}
{Ball}, W.~T., D.~J. {Mortlock}, J.~S. {Egerton}, and J.~D. {Haigh}
  (2014{\natexlab{a}}), {Assessing the relationship between spectral solar
  irradiance and stratospheric ozone using Bayesian inference}, \textit{J.
  Space Weather Space Clim.}, \textit{4}(27), A25, \doi{10.1051/swsc/2014023}.

\bibitem[{\textit{{Ball} et~al.}(2014{\natexlab{b}})\textit{{Ball}, {Krivova},
  {Unruh}, {Haigh}, and {Solanki}}}]{ball14b}
{Ball}, W.~T., N.~A. {Krivova}, Y.~C. {Unruh}, J.~D. {Haigh}, and S.~K.
  {Solanki} (2014{\natexlab{b}}), {A new SATIRE-S spectral solar irradiance
  dataset for solar cycles 21-23 and its implications for stratospheric ozone},
  \textit{J. Atmos. Sci.}, \textit{71}, 4086--4101,
  \doi{10.1175/JAS-D-13-0241.1}.

\bibitem[{\textit{{Balmaceda} et~al.}(2009)\textit{{Balmaceda}, {Solanki},
  {Krivova}, and {Foster}}}]{balmaceda09}
{Balmaceda}, L.~A., S.~K. {Solanki}, N.~A. {Krivova}, and S.~{Foster} (2009),
  {A homogeneous database of sunspot areas covering more than 130 years},
  \textit{J. Geophys. Res.}, \textit{114}, 7104, \doi{10.1029/2009JA014299}.

\bibitem[{\textit{{Bolduc} et~al.}(2012)\textit{{Bolduc}, {Charbonneau},
  {Dumoulin}, {Bourqui}, and {Crouch}}}]{bolduc12}
{Bolduc}, C., P.~{Charbonneau}, V.~{Dumoulin}, M.~S. {Bourqui}, and A.~D.
  {Crouch} (2012), {A Fast Model for the Reconstruction of Spectral Solar
  Irradiance in the Near- and Mid-Ultraviolet}, \textit{Solar Phys.},
  \textit{279}, 383--409, \doi{10.1007/s11207-012-0019-4}.

\bibitem[{\textit{{Brueckner} et~al.}(1993)\textit{{Brueckner}, {Edlow},
  {Floyd}, {Lean}, and {Vanhoosier}}}]{brueckner93}
{Brueckner}, G.~E., K.~L. {Edlow}, L.~E. {Floyd}, IV, J.~L. {Lean}, and M.~E.
  {Vanhoosier} (1993), {The Solar Ultraviolet Spectral Irradiance Monitor
  (SUSIM) experiment on board the Upper Atmosphere Research Satellite (UARS)},
  \textit{J. Geophys. Res.}, \textit{98}, 10,695, \doi{10.1029/93JD00410}.

\bibitem[{\textit{{Cebula} et~al.}(1998)\textit{{Cebula}, {DeLand}, and
  {Hilsenrath}}}]{cebula98}
{Cebula}, R.~P., M.~T. {DeLand}, and E.~{Hilsenrath} (1998), {NOAA 11 solar
  backscattered ultraviolet, model 2 (SBUV/2) instrument solar spectral
  irradiance measurements in 1989-1994 1. Observations and long-term
  calibration}, \textit{J. Geophys. Res.}, \textit{103}, 16,235--16,250,
  \doi{10.1029/98JD01205}.

\bibitem[{\textit{{Cessateur} et~al.}(2015)\textit{{Cessateur}, {Shapiro},
  {Yeo}, {Krivova}, {Tagirov}, {Adams}, and {Schmutz}}}]{cessateur15}
{Cessateur}, G., A.~I. {Shapiro}, K.~L. {Yeo}, N.~A. {Krivova}, R.~{Tagirov},
  W.~{Adams}, and W.~{Schmutz} (2015), {Solar spectral irradiance variations on
  rotation time scales: Comparison between observations and the COSIR model},
  \textit{Astron. Astrophys.}, submitted.

\bibitem[{\textit{{Chapman} et~al.}(2013)\textit{{Chapman}, {Cookson}, and
  {Preminger}}}]{chapman13}
{Chapman}, G.~A., A.~M. {Cookson}, and D.~G. {Preminger} (2013), {Modeling
  Total Solar Irradiance with San Fernando Observatory Ground-Based Photometry:
  Comparison with ACRIM, PMOD, and RMIB Composites}, \textit{Solar Phys.},
  \textit{283}, 295--305, \doi{10.1007/s11207-013-0233-8}.

\bibitem[{\textit{{DeLand} and {Cebula}}(2001)}]{deland01}
{DeLand}, M.~T., and R.~P. {Cebula} (2001), {Spectral solar UV irradiance data
  for cycle 21}, \textit{J. Geophys. Res.}, \textit{106}, 21,569--21,584,
  \doi{10.1029/2000JA000436}.

\bibitem[{\textit{{DeLand} and {Cebula}}(2012)}]{deland12}
{DeLand}, M.~T., and R.~P. {Cebula} (2012), {Solar UV variations during the
  decline of Cycle 23}, \textit{J. Atmos. Sol.-Terr. Phy.}, \textit{77},
  225--234, \doi{10.1016/j.jastp.2012.01.007}.

\bibitem[{\textit{{DeLand} et~al.}(2004)\textit{{DeLand}, {Cebula}, and
  {Hilsenrath}}}]{deland04}
{DeLand}, M.~T., R.~P. {Cebula}, and E.~{Hilsenrath} (2004), {Observations of
  solar spectral irradiance change during cycle 22 from NOAA-9 Solar
  Backscattered Ultraviolet Model 2 (SBUV/2)}, \textit{J. Geophys. Res.},
  \textit{109}, D6304, \doi{10.1029/2003JD004074}.

\bibitem[{\textit{{Dhomse} et~al.}(2013)\textit{{Dhomse}, {Chipperfield},
  {Feng}, {Ball}, {Unruh}, {Haigh}, {Krivova}, {Solanki}, and
  {Smith}}}]{dhomse13}
{Dhomse}, S.~S., M.~P. {Chipperfield}, W.~{Feng}, W.~T. {Ball}, Y.~C. {Unruh},
  J.~D. {Haigh}, N.~A. {Krivova}, S.~K. {Solanki}, and A.~K. {Smith} (2013),
  {Stratospheric O$_{3}$ changes during 2001-2010: the small role of solar flux
  variations in a chemical transport model}, \textit{Atmos. Chem. Phys.},
  \textit{13}, 10,113--10,123, \doi{10.5194/acp-13-10113-2013}.

\bibitem[{\textit{{Domingo} et~al.}(2009)\textit{{Domingo}, {Ermolli}, {Fox},
  {Fr{\"o}hlich}, {Haberreiter}, {Krivova}, {Kopp}, {Schmutz}, {Solanki},
  {Spruit}, {Unruh}, and {V{\"o}gler}}}]{domingo09}
{Domingo}, V., I.~{Ermolli}, P.~{Fox}, C.~{Fr{\"o}hlich}, M.~{Haberreiter},
  N.~{Krivova}, G.~{Kopp}, W.~{Schmutz}, S.~K. {Solanki}, H.~C. {Spruit},
  Y.~{Unruh}, and A.~{V{\"o}gler} (2009), Solar surface magnetism and
  irradiance on time scales from days to the 11-year cycle, \textit{Space Sci.
  Rev.}, \textit{145}, 337--380, \doi{10.1007/s11214-009-9562-1}.

\bibitem[{\textit{{Dudok de Wit} et~al.}(2014)\textit{{Dudok de Wit},
  {Bruinsma}, and {Shibasaki}}}]{dudokdewit14}
{Dudok de Wit}, T., S.~{Bruinsma}, and K.~{Shibasaki} (2014), {Synoptic radio
  observations as proxies for upper atmosphere modelling}, \textit{J. Space
  Weather Space Clim.}, \textit{4}(26), A06, \doi{10.1051/swsc/2014003}.

\bibitem[{\textit{{Ermolli} et~al.}(2003)\textit{{Ermolli}, {Caccin},
  {Centrone}, and {Penza}}}]{ermolli03}
{Ermolli}, I., B.~{Caccin}, M.~{Centrone}, and V.~{Penza} (2003), {Modeling
  solar irradiance variations through full-disk images and semi-empirical
  atmospheric models}, \textit{Mem. Soc. Astron. Italiana}, \textit{74}, 603.

\bibitem[{\textit{{Ermolli} et~al.}(2011)\textit{{Ermolli}, {Criscuoli}, and
  {Giorgi}}}]{ermolli11}
{Ermolli}, I., S.~{Criscuoli}, and F.~{Giorgi} (2011), {Recent results from
  optical synoptic observations of the solar atmosphere with ground-based
  instruments}, \textit{Cont. Astron. Obs. Skalnat{\'e} Pleso}, \textit{41},
  73--84.

\bibitem[{\textit{{Ermolli} et~al.}(2013)\textit{{Ermolli}, {Matthes}, {Dudok
  de Wit}, {Krivova}, {Tourpali}, {Weber}, {Unruh}, {Gray}, {Langematz},
  {Pilewskie}, {Rozanov}, {Schmutz}, {Shapiro}, {Solanki}, and
  {Woods}}}]{ermolli13}
{Ermolli}, I., K.~{Matthes}, T.~{Dudok de Wit}, N.~A. {Krivova}, K.~{Tourpali},
  M.~{Weber}, Y.~C. {Unruh}, L.~{Gray}, U.~{Langematz}, P.~{Pilewskie},
  E.~{Rozanov}, W.~{Schmutz}, A.~{Shapiro}, S.~K. {Solanki}, and T.~N. {Woods}
  (2013), {Recent variability of the solar spectral irradiance and its impact
  on climate modelling}, \textit{Atmos. Chem. Phys.}, \textit{13}, 3945--3977,
  \doi{10.5194/acp-13-3945-2013}.

\bibitem[{\textit{{Ermolli} et~al.}(2014)\textit{{Ermolli}, {Shibasaki},
  {Tlatov}, and {van Driel-Gesztelyi}}}]{ermolli14}
{Ermolli}, I., K.~{Shibasaki}, A.~{Tlatov}, and L.~{van Driel-Gesztelyi}
  (2014), {Solar Cycle Indices from the Photosphere to the Corona: Measurements
  and Underlying Physics}, \textit{Space Sci. Rev.},
  \doi{10.1007/s11214-014-0089-8}.

\bibitem[{\textit{{Fligge} et~al.}(2000)\textit{{Fligge}, {Solanki}, and
  {Unruh}}}]{fligge00}
{Fligge}, M., S.~K. {Solanki}, and Y.~C. {Unruh} (2000), {Modelling irradiance
  variations from the surface distribution of the solar magnetic field},
  \textit{Astron. Astrophys.}, \textit{353}, 380--388.

\bibitem[{\textit{{Floyd} et~al.}(2003)\textit{{Floyd}, {Cook}, {Herring}, and
  {Crane}}}]{floyd03}
{Floyd}, L.~E., J.~W. {Cook}, L.~C. {Herring}, and P.~C. {Crane} (2003),
  {SUSIM'S 11-year observational record of the solar UV irradiance},
  \textit{Adv. Space Res.}, \textit{31}, 2111--2120,
  \doi{10.1016/S0273-1177(03)00148-0}.

\bibitem[{\textit{{Fontenla} et~al.}(1999)\textit{{Fontenla}, {White}, {Fox},
  {Avrett}, and {Kurucz}}}]{fontenla99}
{Fontenla}, J., O.~R. {White}, P.~A. {Fox}, E.~H. {Avrett}, and R.~L. {Kurucz}
  (1999), {Calculation of solar irradiances. I. Synthesis of the solar
  spectrum}, \textit{Astrophys. J.}, \textit{518}, 480--499,
  \doi{10.1086/307258}.

\bibitem[{\textit{{Fontenla} et~al.}(2004)\textit{{Fontenla}, {Harder},
  {Rottman}, {Woods}, {Lawrence}, and {Davis}}}]{fontenla04}
{Fontenla}, J.~M., J.~{Harder}, G.~{Rottman}, T.~N. {Woods}, G.~M. {Lawrence},
  and S.~{Davis} (2004), The signature of solar activity in the infrared
  spectral irradiance, \textit{Astrophys. J.}, \textit{605}, L85--L88,
  \doi{10.1086/386335}.

\bibitem[{\textit{{Fontenla} et~al.}(2006)\textit{{Fontenla}, {Avrett},
  {Thuillier}, and {Harder}}}]{fontenla06}
{Fontenla}, J.~M., E.~{Avrett}, G.~{Thuillier}, and J.~{Harder} (2006),
  {Semiempirical models of the solar atmosphere. I. The quiet- and active Sun
  photosphere at moderate resolution}, \textit{Astrophys. J.}, \textit{639},
  441--458, \doi{10.1086/499345}.

\bibitem[{\textit{{Fontenla} et~al.}(2009)\textit{{Fontenla}, {Curdt},
  {Haberreiter}, {Harder}, and {Tian}}}]{fontenla09}
{Fontenla}, J.~M., W.~{Curdt}, M.~{Haberreiter}, J.~{Harder}, and H.~{Tian}
  (2009), {Semiempirical models of the solar atmosphere. III. Set of non-LTE
  models for far-ultraviolet/extreme-ultraviolet irradiance computation},
  \textit{Astrophys. J.}, \textit{707}, 482--502,
  \doi{10.1088/0004-637X/707/1/482}.

\bibitem[{\textit{{Fontenla} et~al.}(2011)\textit{{Fontenla}, {Harder},
  {Livingston}, {Snow}, and {Woods}}}]{fontenla11}
{Fontenla}, J.~M., J.~{Harder}, W.~{Livingston}, M.~{Snow}, and T.~{Woods}
  (2011), {High-resolution solar spectral irradiance from extreme ultraviolet
  to far infrared}, \textit{J. Geophys. Res.}, \textit{116}, D20,108,
  \doi{10.1029/2011JD016032}.

\bibitem[{\textit{{Foukal} and {Lean}}(1986)}]{foukal86}
{Foukal}, P., and J.~{Lean} (1986), {The influence of faculae on total solar
  irradiance and luminosity}, \textit{Astrophys. J.}, \textit{302}, 826--835,
  \doi{10.1086/164043}.

\bibitem[{\textit{{Fr{\"o}hlich}}(2012)}]{frohlich12}
{Fr{\"o}hlich}, C. (2012), {Total solar irradiance observations}, \textit{Surv.
  Geophys.}, \textit{33}, 453--473, \doi{10.1007/s10712-011-9168-5}.

\bibitem[{\textit{{Fr{\"o}hlich}}(2013)}]{frohlich13}
{Fr{\"o}hlich}, C. (2013), Total solar irradiance: What have we learned from
  the last three cycles and the recent minimum?, \textit{Space Sci. Rev.},
  \textit{176}, 237--252, \doi{10.1007/s11214-011-9780-1}.

\bibitem[{\textit{{Fr{\"o}hlich} and {Lean}}(2004)}]{frohlich04}
{Fr{\"o}hlich}, C., and J.~{Lean} (2004), {Solar radiative output and its
  variability: evidence and mechanisms}, \textit{Astron. Astrophy. Rev.},
  \textit{12}, 273--320, \doi{10.1007/s00159-004-0024-1}.

\bibitem[{\textit{{Fr{\"o}hlich} et~al.}(1994)\textit{{Fr{\"o}hlich}, {Pap},
  and {Hudson}}}]{frohlich94}
{Fr{\"o}hlich}, C., J.~M. {Pap}, and H.~S. {Hudson} (1994), {Improvement of the
  photometric sunspot index and changes of the disk-integrated sunspot contrast
  with time}, \textit{Solar Phys.}, \textit{152}, 111--118,
  \doi{10.1007/BF01473192}.

\bibitem[{\textit{{Gray} et~al.}(2010)\textit{{Gray}, {Beer}, {Geller},
  {Haigh}, {Lockwood}, {Matthes}, {Cubasch}, {Fleitmann}, {Harrison}, {Hood},
  {Luterbacher}, {Meehl}, {Shindell}, {van Geel}, and {White}}}]{gray10}
{Gray}, L.~J., J.~{Beer}, M.~{Geller}, J.~D. {Haigh}, M.~{Lockwood},
  K.~{Matthes}, U.~{Cubasch}, D.~{Fleitmann}, G.~{Harrison}, L.~{Hood},
  J.~{Luterbacher}, G.~A. {Meehl}, D.~{Shindell}, B.~{van Geel}, and W.~{White}
  (2010), Solar influences on climate, \textit{Rev. Geophys.}, \textit{48}(4),
  RG4001, \doi{10.1029/2009RG000282}.

\bibitem[{\textit{{Haberreiter} et~al.}(2005)\textit{{Haberreiter}, {Krivova},
  {Schmutz}, and {Wenzler}}}]{haberreiter05}
{Haberreiter}, M., N.~A. {Krivova}, W.~{Schmutz}, and T.~{Wenzler} (2005),
  {Reconstruction of the solar UV irradiance back to 1974}, \textit{Adv. Space
  Res.}, \textit{35}, 365--369, \doi{10.1016/j.asr.2005.04.039}.

\bibitem[{\textit{{Haberreiter} et~al.}(2008)\textit{{Haberreiter}, {Schmutz},
  and {Hubeny}}}]{haberreiter08}
{Haberreiter}, M., W.~{Schmutz}, and I.~{Hubeny} (2008), {NLTE model
  calculations for the solar atmosphere with an iterative treatment of opacity
  distribution functions}, \textit{Astron. Astrophys.}, \textit{492}, 833--840,
  \doi{10.1051/0004-6361:200809503}.

\bibitem[{\textit{{Haigh}}(2007)}]{haigh07}
{Haigh}, J.~D. (2007), {The Sun and the Earth's Climate}, \textit{Living Rev.
  Solar Phys.}, \textit{4}, 2, \doi{10.12942/lrsp-2007-2}.

\bibitem[{\textit{{Haigh} et~al.}(2010)\textit{{Haigh}, {Winning}, {Toumi}, and
  {Harder}}}]{haigh10}
{Haigh}, J.~D., A.~R. {Winning}, R.~{Toumi}, and J.~W. {Harder} (2010), {An
  influence of solar spectral variations on radiative forcing of climate},
  \textit{Nature}, \textit{467}(7316), 696--699, \doi{10.1038/nature09426}.

\bibitem[{\textit{{Harder} et~al.}(2005{\natexlab{a}})\textit{{Harder},
  {Lawrence}, {Fontenla}, {Rottman}, and {Woods}}}]{harder05a}
{Harder}, J., G.~{Lawrence}, J.~{Fontenla}, G.~{Rottman}, and T.~{Woods}
  (2005{\natexlab{a}}), {The Spectral Irradiance Monitor: scientific
  requirements, instrument design, and operation modes}, \textit{Solar Phys.},
  \textit{230}, 141--167, \doi{10.1007/s11207-005-5007-5}.

\bibitem[{\textit{{Harder} et~al.}(2005{\natexlab{b}})\textit{{Harder},
  {Fontenla}, {Lawrence}, {Woods}, and {Rottman}}}]{harder05b}
{Harder}, J.~W., J.~{Fontenla}, G.~{Lawrence}, T.~{Woods}, and G.~{Rottman}
  (2005{\natexlab{b}}), {The Spectral Irradiance Monitor: measurement equations
  and calibration}, \textit{Solar Phys.}, \textit{230}, 169--204,
  \doi{10.1007/s11207-005-1528-1}.

\bibitem[{\textit{{Hickey} et~al.}(1988)\textit{{Hickey}, {Alton}, {Kyle}, and
  {Hoyt}}}]{hickey88}
{Hickey}, J.~R., B.~M. {Alton}, H.~L. {Kyle}, and D.~{Hoyt} (1988), {Total
  solar irradiance measurements by ERB/Nimbus-7 - A review of nine years},
  \textit{Space Sci. Rev.}, \textit{48}, 321--342, \doi{10.1007/BF00226011}.

\bibitem[{\textit{{Holzreuter} and {Solanki}}(2013)}]{holzreuter13}
{Holzreuter}, R., and S.~K. {Solanki} (2013), {Three-dimensional non-LTE
  radiative transfer effects in Fe I lines. II. Line formation in 3D radiation
  hydrodynamic simulations}, \textit{Astron. Astrophys.}, \textit{558}, A20,
  \doi{10.1051/0004-6361/201322135}.

\bibitem[{\textit{{Hudson}}(1988)}]{hudson88}
{Hudson}, H.~S. (1988), {Observed variability of the solar luminosity},
  \textit{Annu. Rev. Astro. Astrophys.}, \textit{26}, 473--507,
  \doi{10.1146/annurev.aa.26.090188.002353}.

\bibitem[{\textit{{Hudson} et~al.}(1982)\textit{{Hudson}, {Silva}, {Woodard},
  and {Willson}}}]{hudson82}
{Hudson}, H.~S., S.~{Silva}, M.~{Woodard}, and R.~C. {Willson} (1982), {The
  effects of sunspots on solar irradiance}, \textit{Solar Phys.}, \textit{76},
  211--219, \doi{10.1007/BF00170984}.

\bibitem[{\textit{{Jones} et~al.}(1992)\textit{{Jones}, {Duvall}, {Harvey},
  {Mahaffey}, {Schwitters}, and {Simmons}}}]{jones92}
{Jones}, H.~P., T.~L. {Duvall}, Jr., J.~W. {Harvey}, C.~T. {Mahaffey}, J.~D.
  {Schwitters}, and J.~E. {Simmons} (1992), {The NASA/NSO spectromagnetograph},
  \textit{Solar Phys.}, \textit{139}, 211--232, \doi{10.1007/BF00159149}.

\bibitem[{\textit{{Kopp}}(2014)}]{kopp14}
{Kopp}, G. (2014), An assessment of the solar irradiance record for climate
  studies, \textit{J. Space Weather Space Clim.}, \textit{4}, A14,
  \doi{10.1051/swsc/2014012}.

\bibitem[{\textit{{Krivova} et~al.}(2003)\textit{{Krivova}, {Solanki},
  {Fligge}, and {Unruh}}}]{krivova03}
{Krivova}, N.~A., S.~K. {Solanki}, M.~{Fligge}, and Y.~C. {Unruh} (2003),
  {Reconstruction of solar irradiance variations in cycle 23: Is solar surface
  magnetism the cause?}, \textit{Astron. Astrophys.}, \textit{399}, L1--L4,
  \doi{10.1051/0004-6361:20030029}.

\bibitem[{\textit{{Krivova} et~al.}(2006)\textit{{Krivova}, {Solanki}, and
  {Floyd}}}]{krivova06}
{Krivova}, N.~A., S.~K. {Solanki}, and L.~{Floyd} (2006), {Reconstruction of
  solar UV irradiance in cycle 23}, \textit{Astron. Astrophys.}, \textit{452},
  631--639, \doi{10.1051/0004-6361:20064809}.

\bibitem[{\textit{{Krivova} et~al.}(2009)\textit{{Krivova}, {Solanki}, and
  {Wenzler}}}]{krivova09}
{Krivova}, N.~A., S.~K. {Solanki}, and T.~{Wenzler} (2009), {ACRIM-gap and
  total solar irradiance revisited: Is there a secular trend between 1986 and
  1996?}, \textit{Geophys. Res. Lett.}, \textit{36}, 20,101,
  \doi{10.1029/2009GL040707}.

\bibitem[{\textit{{Kurucz}}(1991)}]{kurucz91}
{Kurucz}, R.~L. (1991), {New Opacity Calculations}, in \textit{NATO Advanced
  Science Institutes (ASI) Series C}, \textit{NATO Advanced Science Institutes
  (ASI) Series C}, vol. 341, edited by L.~{Crivellari}, I.~{Hubeny}, and D.~G.
  {Hummer}, p. 441.

\bibitem[{\textit{{Lean}}(2000)}]{lean00}
{Lean}, J. (2000), {Evolution of the Sun's spectral irradiance since the
  Maunder minimum}, \textit{Geophys. Res. Lett.}, \textit{27}, 2425--2428,
  \doi{10.1029/2000GL000043}.

\bibitem[{\textit{{Lean} and {DeLand}}(2012)}]{lean12}
{Lean}, J.~L., and M.~T. {DeLand} (2012), {How does the Sun's spectrum vary?},
  \textit{J. Climate}, \textit{25}, 2555--2560,
  \doi{10.1175/JCLI-D-11-00571.1}.

\bibitem[{\textit{{Lean} et~al.}(1997)\textit{{Lean}, {Rottman}, {Kyle},
  {Woods}, {Hickey}, and {Puga}}}]{lean97}
{Lean}, J.~L., G.~J. {Rottman}, H.~L. {Kyle}, T.~N. {Woods}, J.~R. {Hickey},
  and L.~C. {Puga} (1997), {Detection and parameterization of variations in
  solar mid- and near-ultraviolet radiation (200-400 nm)}, \textit{J. Geophys.
  Res.}, \textit{102}, 29,939--29,956, \doi{10.1029/97JD02092}.

\bibitem[{\textit{{Lean} et~al.}(1998)\textit{{Lean}, {Cook}, {Marquette}, and
  {Johannesson}}}]{lean98}
{Lean}, J.~L., J.~{Cook}, W.~{Marquette}, and A.~{Johannesson} (1998),
  {Magnetic Sources of the Solar Irradiance Cycle}, \textit{Astrophys. J.},
  \textit{492}, 390--401, \doi{10.1086/305015}.

\bibitem[{\textit{{Lilensten} et~al.}(2008)\textit{{Lilensten}, {Dudok de Wit},
  {Kretzschmar}, {Amblard}, {Moussaoui}, {Aboudarham}, and
  {Auch{\`e}re}}}]{lilensten08}
{Lilensten}, J., T.~{Dudok de Wit}, M.~{Kretzschmar}, P.-O. {Amblard},
  S.~{Moussaoui}, J.~{Aboudarham}, and F.~{Auch{\`e}re} (2008), {Review on the
  solar spectral variability in the EUV for space weather purposes},
  \textit{Annales Geophysicae}, \textit{26}, 269--279,
  \doi{10.5194/angeo-26-269-2008}.

\bibitem[{\textit{{Livingston} et~al.}(1976)\textit{{Livingston}, {Harvey},
  {Pierce}, {Schrage}, {Gillespie}, {Simmons}, and {Slaughter}}}]{livingston76}
{Livingston}, W.~C., J.~{Harvey}, A.~K. {Pierce}, D.~{Schrage}, B.~{Gillespie},
  J.~{Simmons}, and C.~{Slaughter} (1976), {Kitt Peak 60-cm vacuum telescope},
  \textit{Appl. Opt.}, \textit{15}, 33--39, \doi{10.1364/AO.15.000033}.

\bibitem[{\textit{{Marchenko} and {DeLand}}(2014)}]{marchenko14}
{Marchenko}, S.~V., and M.~T. {DeLand} (2014), {Solar Spectral Irradiance
  Changes during Cycle 24}, \textit{Astrophys. J.}, \textit{789}, 117,
  \doi{10.1088/0004-637X/789/2/117}.

\bibitem[{\textit{{McClintock} et~al.}(2005)\textit{{McClintock}, {Rottman},
  and {Woods}}}]{mcclintock05}
{McClintock}, W.~E., G.~J. {Rottman}, and T.~N. {Woods} (2005), {Solar-Stellar
  Irradiance Comparison Experiment II (Solstice II): Instrument Concept and
  Design}, \textit{Solar Phys.}, \textit{230}, 225--258,
  \doi{10.1007/s11207-005-7432-x}.

\bibitem[{\textit{{Merkel} et~al.}(2011)\textit{{Merkel}, {Harder}, {Marsh},
  {Smith}, {Fontenla}, and {Woods}}}]{merkel11}
{Merkel}, A.~W., J.~W. {Harder}, D.~R. {Marsh}, A.~K. {Smith}, J.~M.
  {Fontenla}, and T.~N. {Woods} (2011), {The impact of solar spectral
  irradiance variability on middle atmospheric ozone}, \textit{Geophys. Res.
  Lett.}, \textit{38}, L13,802, \doi{10.1029/2011GL047561}.

\bibitem[{\textit{{Morrill} et~al.}(2011)\textit{{Morrill}, {Floyd}, and
  {McMullin}}}]{morrill11}
{Morrill}, J.~S., L.~{Floyd}, and D.~{McMullin} (2011), {The solar ultraviolet
  spectrum estimated using the Mg II index and Ca II K disk activity},
  \textit{Solar Phys.}, \textit{269}, 253--267,
  \doi{10.1007/s11207-011-9708-7}.

\bibitem[{\textit{{Morrill} et~al.}(2014)\textit{{Morrill}, {Floyd}, and
  {McMullin}}}]{morrill14}
{Morrill}, J.~S., L.~{Floyd}, and D.~{McMullin} (2014), {Comparison of Solar UV
  Spectral Irradiance from SUSIM and SORCE}, \textit{Solar Phys.},
  \textit{289}, 3641--3661, \doi{10.1007/s11207-014-0535-5}.

\bibitem[{\textit{{Oberl{\"a}nder} et~al.}(2012)\textit{{Oberl{\"a}nder},
  {Langematz}, {Matthes}, {Kunze}, {Kubin}, {Harder}, {Krivova}, {Solanki},
  {Pagaran}, and {Weber}}}]{oberlander12}
{Oberl{\"a}nder}, S., U.~{Langematz}, K.~{Matthes}, M.~{Kunze}, A.~{Kubin},
  J.~{Harder}, N.~A. {Krivova}, S.~K. {Solanki}, J.~{Pagaran}, and M.~{Weber}
  (2012), {The influence of spectral solar irradiance data on stratospheric
  heating rates during the 11 year solar cycle}, \textit{Geophys. Res. Lett.},
  \textit{39}, L01801, \doi{10.1029/2011GL049539}.

\bibitem[{\textit{{Oster} et~al.}(1982)\textit{{Oster}, {Schatten}, and
  {Sofia}}}]{oster82}
{Oster}, L., K.~H. {Schatten}, and S.~{Sofia} (1982), {Solar irradiance
  variations due to active regions}, \textit{Astrophys. J.}, \textit{256},
  768--773, \doi{10.1086/159949}.

\bibitem[{\textit{{Penza} et~al.}(2003)\textit{{Penza}, {Caccin}, {Ermolli},
  {Centrone}, and {Gomez}}}]{penza03}
{Penza}, V., B.~{Caccin}, I.~{Ermolli}, M.~{Centrone}, and M.~T. {Gomez}
  (2003), {Modeling solar irradiance variations through PSPT images and
  semiempirical models}, in \textit{Solar Variability as an Input to the
  Earth's Environment}, \textit{ESA Sp. Pub.}, vol. 535, edited by A.~{Wilson},
  pp. 299--302.

\bibitem[{\textit{{Rottman}}(2005)}]{rottman05}
{Rottman}, G. (2005), {The SORCE mission}, \textit{Solar Phys.}, \textit{230},
  7--25, \doi{10.1007/s11207-005-8112-6}.

\bibitem[{\textit{{Rottman} et~al.}(2001)\textit{{Rottman}, {Woods}, {Snow},
  and {DeToma}}}]{rottman01}
{Rottman}, G., T.~{Woods}, M.~{Snow}, and G.~{DeToma} (2001), {The solar cycle
  variation in ultraviolet irradiance}, \textit{Adv. Space Res.}, \textit{27},
  1927--1932, \doi{10.1016/S0273-1177(01)00272-1}.

\bibitem[{\textit{{Rottman}}(1988)}]{rottman88}
{Rottman}, G.~J. (1988), {Observations of solar UV and EUV variability},
  \textit{Adv. Space Res.}, \textit{8}, 53--66,
  \doi{10.1016/0273-1177(88)90172-X}.

\bibitem[{\textit{{Scafetta} and {Willson}}(2009)}]{scafetta09}
{Scafetta}, N., and R.~C. {Willson} (2009), {ACRIM-gap and TSI trend issue
  resolved using a surface magnetic flux TSI proxy model}, \textit{Geophys.
  Res. Lett.}, \textit{36}, 5701, \doi{10.1029/2008GL036307}.

\bibitem[{\textit{{Scherrer} et~al.}(1995)\textit{{Scherrer}, {Bogart}, {Bush},
  {Hoeksema}, {Kosovichev}, {Schou}, {Rosenberg}, {Springer}, {Tarbell},
  {Title}, {Wolfson}, {Zayer}, and {MDI Engineering Team}}}]{scherrer95}
{Scherrer}, P.~H., R.~S. {Bogart}, R.~I. {Bush}, J.~T. {Hoeksema}, A.~G.
  {Kosovichev}, J.~{Schou}, W.~{Rosenberg}, L.~{Springer}, T.~D. {Tarbell},
  A.~{Title}, C.~J. {Wolfson}, I.~{Zayer}, and {MDI Engineering Team} (1995),
  {The Solar Oscillations Investigation - Michelson Doppler Imager},
  \textit{Solar Phys.}, \textit{162}, 129--188, \doi{10.1007/BF00733429}.

\bibitem[{\textit{{Schou} et~al.}(2012)\textit{{Schou}, {Scherrer}, {Bush},
  {Wachter}, {Couvidat}, {Rabello-Soares}, {Bogart}, {Hoeksema}, {Liu},
  {Duvall}, {Akin}, {Allard}, {Miles}, {Rairden}, {Shine}, {Tarbell}, {Title},
  {Wolfson}, {Elmore}, {Norton}, and {Tomczyk}}}]{schou12}
{Schou}, J., P.~H. {Scherrer}, R.~I. {Bush}, R.~{Wachter}, S.~{Couvidat}, M.~C.
  {Rabello-Soares}, R.~S. {Bogart}, J.~T. {Hoeksema}, Y.~{Liu}, T.~L. {Duvall},
  D.~J. {Akin}, B.~A. {Allard}, J.~W. {Miles}, R.~{Rairden}, R.~A. {Shine},
  T.~D. {Tarbell}, A.~M. {Title}, C.~J. {Wolfson}, D.~F. {Elmore}, A.~A.
  {Norton}, and S.~{Tomczyk} (2012), {Design and Ground Calibration of the
  Helioseismic and Magnetic Imager (HMI) Instrument on the Solar Dynamics
  Observatory (SDO)}, \textit{Solar Phys.}, \textit{275}, 229--259,
  \doi{10.1007/s11207-011-9842-2}.

\bibitem[{\textit{{Schwenn}}(2006)}]{schwenn06}
{Schwenn}, R. (2006), {Space Weather: The Solar Perspective}, \textit{Living
  Rev. Solar Phys.}, \textit{3}, 2, \doi{10.12942/lrsp-2006-2}.

\bibitem[{\textit{{Seleznyov} et~al.}(2011)\textit{{Seleznyov}, {Solanki}, and
  {Krivova}}}]{seleznyov11}
{Seleznyov}, A.~D., S.~K. {Solanki}, and N.~A. {Krivova} (2011), {Modelling
  solar irradiance variability on time scales from minutes to months},
  \textit{Astron. Astrophys.}, \textit{532}, A108,
  \doi{10.1051/0004-6361/200811138}.

\bibitem[{\textit{{Shapiro} et~al.}(2010)\textit{{Shapiro}, {Schmutz},
  {Schoell}, {Haberreiter}, and {Rozanov}}}]{shapiro10}
{Shapiro}, A.~I., W.~{Schmutz}, M.~{Schoell}, M.~{Haberreiter}, and
  E.~{Rozanov} (2010), {NLTE solar irradiance modeling with the COSI code},
  \textit{Astron. Astrophys.}, \textit{517}, A48,
  \doi{10.1051/0004-6361/200913987}.

\bibitem[{\textit{{Shapiro} et~al.}(2011)\textit{{Shapiro}, {Schmutz},
  {Rozanov}, {Schoell}, {Haberreiter}, {Shapiro}, and {Nyeki}}}]{shapiro11}
{Shapiro}, A.~I., W.~{Schmutz}, E.~{Rozanov}, M.~{Schoell}, M.~{Haberreiter},
  A.~V. {Shapiro}, and S.~{Nyeki} (2011), {A new approach to the long-term
  reconstruction of the solar irradiance leads to large historical solar
  forcing}, \textit{Astron. Astrophys.}, \textit{529}, A67,
  \doi{10.1051/0004-6361/201016173}.

\bibitem[{\textit{{Shapiro} et~al.}(2013)\textit{{Shapiro}, {Shapiro},
  {Dominique}, {Dammasch}, {Wehrli}, {Rozanov}, and {Schmutz}}}]{shapiro13}
{Shapiro}, A.~V., A.~I. {Shapiro}, M.~{Dominique}, I.~E. {Dammasch},
  C.~{Wehrli}, E.~{Rozanov}, and W.~{Schmutz} (2013), {Detection of solar
  rotational variability in the Large Yield RAdiometer (LYRA) 190 - 222 nm
  spectral band}, \textit{Solar Phys.}, \textit{286}, 289--301,
  \doi{10.1007/s11207-012-0029-2}.

\bibitem[{\textit{{Snow} et~al.}(2005{\natexlab{a}})\textit{{Snow},
  {McClintock}, {Rottman}, and {Woods}}}]{snow05a}
{Snow}, M., W.~E. {McClintock}, G.~{Rottman}, and T.~N. {Woods}
  (2005{\natexlab{a}}), {Solar Stellar Irradiance Comparison Experiment II
  (Solstice II): Examination of the Solar Stellar Comparison Technique},
  \textit{Solar Phys.}, \textit{230}, 295--324,
  \doi{10.1007/s11207-005-8763-3}.

\bibitem[{\textit{{Snow} et~al.}(2005{\natexlab{b}})\textit{{Snow},
  {McClintock}, {Woods}, {White}, {Harder}, and {Rottman}}}]{snow05b}
{Snow}, M., W.~E. {McClintock}, T.~N. {Woods}, O.~R. {White}, J.~W. {Harder},
  and G.~{Rottman} (2005{\natexlab{b}}), {The Mg II Index from SORCE},
  \textit{Solar Phys.}, \textit{230}, 325--344,
  \doi{10.1007/s11207-005-6879-0}.

\bibitem[{\textit{{Solanki} and {Unruh}}(1998)}]{solanki98}
{Solanki}, S.~K., and Y.~C. {Unruh} (1998), {A model of the wavelength
  dependence of solar irradiance variations}, \textit{Astron. Astrophys.},
  \textit{329}, 747--753.

\bibitem[{\textit{{Solanki} et~al.}(2013)\textit{{Solanki}, {Krivova}, and
  {Haigh}}}]{solanki13}
{Solanki}, S.~K., N.~A. {Krivova}, and J.~D. {Haigh} (2013), Solar irradiance
  variability and climate, \textit{Annu. Rev. Astron. Astrophys.}, \textit{51},
  311--351, \doi{10.1146/annurev-astro-082812-141007}.

\bibitem[{\textit{{Tapping}}(2013)}]{tapping13}
{Tapping}, K.~F. (2013), {The 10.7 cm solar radio flux (F10.7)}, \textit{Adv.
  Space Res.}, \textit{11}(7), 394--406, \doi{10.1002/swe.20064}.

\bibitem[{\textit{{Thuillier} et~al.}(1998)\textit{{Thuillier}, {Herse},
  {Simon}, {Labs}, {Mandel}, {Gillotay}, and {Foujols}}}]{thuillier98}
{Thuillier}, G., M.~{Herse}, P.~C. {Simon}, D.~{Labs}, H.~{Mandel},
  D.~{Gillotay}, and T.~{Foujols} (1998), {The Visible Solar Spectral
  Irradiance from 350 to 850 NM as Measured by the SOLSPEC Spectrometer During
  the Atlas I Mission}, \textit{Solar Phys.}, \textit{177}, 41--61,
  \doi{10.1023/A:1004953215589}.

\bibitem[{\textit{{Thuillier} et~al.}(2012)\textit{{Thuillier}, {Deland},
  {Shapiro}, {Schmutz}, {Bols{\'e}e}, and {Melo}}}]{thuillier12}
{Thuillier}, G., M.~{Deland}, A.~{Shapiro}, W.~{Schmutz}, D.~{Bols{\'e}e}, and
  S.~M.~L. {Melo} (2012), {The solar spectral irradiance as a function of the
  Mg II index for atmosphere and climate modelling}, \textit{Solar Phys.},
  \textit{277}, 245--266, \doi{10.1007/s11207-011-9912-5}.

\bibitem[{\textit{{Uitenbroek} and {Criscuoli}}(2011)}]{uitenbroek11}
{Uitenbroek}, H., and S.~{Criscuoli} (2011), Why one-dimensional models fail in
  the diagnosis of average spectra from inhomogeneous stellar atmospheres,
  \textit{Astrophys. J.}, \textit{736}, 69, \doi{10.1088/0004-637X/736/1/69}.

\bibitem[{\textit{{Unruh} et~al.}(1999)\textit{{Unruh}, {Solanki}, and
  {Fligge}}}]{unruh99}
{Unruh}, Y.~C., S.~K. {Solanki}, and M.~{Fligge} (1999), {The spectral
  dependence of facular contrast and solar irradiance variations},
  \textit{Astron. Astrophys.}, \textit{345}, 635--642.

\bibitem[{\textit{{Unruh} et~al.}(2012)\textit{{Unruh}, {Ball}, and
  {Krivova}}}]{unruh12}
{Unruh}, Y.~C., W.~T. {Ball}, and N.~A. {Krivova} (2012), {Solar irradiance
  models and measurements: A comparison in the 220-240 nm wavelength band},
  \textit{Surv. Geophys.}, \textit{33}, 475--481,
  \doi{10.1007/s10712-011-9166-7}.

\bibitem[{\textit{{Viereck} et~al.}(2004)\textit{{Viereck}, {Floyd}, {Crane},
  {Woods}, {Knapp}, {Rottman}, {Weber}, {Puga}, and {Deland}}}]{viereck04}
{Viereck}, R.~A., L.~E. {Floyd}, P.~C. {Crane}, T.~N. {Woods}, B.~G. {Knapp},
  G.~{Rottman}, M.~{Weber}, L.~C. {Puga}, and M.~T. {Deland} (2004), {A
  composite Mg II index spanning from 1978 to 2003}, \textit{Adv. Space Res.},
  \textit{2}, 10,005, \doi{10.1029/2004SW000084}.

\bibitem[{\textit{{Wehrli} et~al.}(2013)\textit{{Wehrli}, {Schmutz}, and
  {Shapiro}}}]{wehrli13}
{Wehrli}, C., W.~{Schmutz}, and A.~I. {Shapiro} (2013), {Correlation of
  spectral solar irradiance with solar activity as measured by VIRGO},
  \textit{Astron. Astrophys.}, \textit{556}, L3,
  \doi{10.1051/0004-6361/201220864}.

\bibitem[{\textit{{Wenzler} et~al.}(2006)\textit{{Wenzler}, {Solanki},
  {Krivova}, and {Fr{\"o}hlich}}}]{wenzler06}
{Wenzler}, T., S.~K. {Solanki}, N.~A. {Krivova}, and C.~{Fr{\"o}hlich} (2006),
  {Reconstruction of solar irradiance variations in cycles 21-23 based on
  surface magnetic fields}, \textit{Astron. Astrophys.}, \textit{460},
  583--595, \doi{10.1051/0004-6361:20065752}.

\bibitem[{\textit{{Willson} and {Hudson}}(1988)}]{willson88}
{Willson}, R.~C., and H.~S. {Hudson} (1988), {Solar luminosity variations in
  solar cycle 21}, \textit{Nature}, \textit{332}, 810--812,
  \doi{10.1038/332810a0}.

\bibitem[{\textit{{Willson} et~al.}(1981)\textit{{Willson}, {Gulkis},
  {Janssen}, {Hudson}, and {Chapman}}}]{willson81}
{Willson}, R.~C., S.~{Gulkis}, M.~{Janssen}, H.~S. {Hudson}, and G.~A.
  {Chapman} (1981), {Observations of solar irradiance variability},
  \textit{Science}, \textit{211}, 700--702, \doi{10.1126/science.211.4483.700}.

\bibitem[{\textit{{Woods}}(2008)}]{woods08}
{Woods}, T.~N. (2008), {Recent advances in observations and modeling of the
  solar ultraviolet and X-ray spectral irradiance}, \textit{Adv. Space Res.},
  \textit{42}, 895--902, \doi{10.1016/j.asr.2007.09.026}.

\bibitem[{\textit{{Woods} et~al.}(1996)\textit{{Woods}, {Prinz}, {Rottman},
  {London}, {Crane}, {Cebula}, {Hilsenrath}, {Brueckner}, {Andrews}, {White},
  {VanHoosier}, {Floyd}, {Herring}, {Knapp}, {Pankratz}, and
  {Reiser}}}]{woods96}
{Woods}, T.~N., D.~K. {Prinz}, G.~J. {Rottman}, J.~{London}, P.~C. {Crane},
  R.~P. {Cebula}, E.~{Hilsenrath}, G.~E. {Brueckner}, M.~D. {Andrews}, O.~R.
  {White}, M.~E. {VanHoosier}, L.~E. {Floyd}, L.~C. {Herring}, B.~G. {Knapp},
  C.~K. {Pankratz}, and P.~A. {Reiser} (1996), {Validation of the UARS solar
  ultraviolet irradiances: Comparison with the ATLAS 1 and 2 measurements},
  \textit{J. Geophys. Res.}, \textit{101}, 9541--9570, \doi{10.1029/96JD00225}.

\bibitem[{\textit{{Woods} et~al.}(2005)\textit{{Woods}, {Eparvier}, {Bailey},
  {Chamberlin}, {Lean}, {Rottman}, {Solomon}, {Tobiska}, and
  {Woodraska}}}]{woods05}
{Woods}, T.~N., F.~G. {Eparvier}, S.~M. {Bailey}, P.~C. {Chamberlin},
  J.~{Lean}, G.~J. {Rottman}, S.~C. {Solomon}, W.~K. {Tobiska}, and D.~L.
  {Woodraska} (2005), {Solar EUV Experiment (SEE): Mission overview and first
  results}, \textit{J. Geophys. Res.}, \textit{110}, A1312,
  \doi{10.1029/2004JA010765}.

\bibitem[{\textit{{Woods} et~al.}(2006)\textit{{Woods}, {Kopp}, and
  {Chamberlin}}}]{woods06}
{Woods}, T.~N., G.~{Kopp}, and P.~C. {Chamberlin} (2006), {Contributions of the
  solar ultraviolet irradiance to the total solar irradiance during large
  flares}, \textit{J. Geophys. Res.}, \textit{111}(A10), 10,
  \doi{10.1029/2005JA011507}.

\bibitem[{\textit{{Woods} et~al.}(2009)\textit{{Woods}, {Chamberlin}, {Harder},
  {Hock}, {Snow}, {Eparvier}, {Fontenla}, {McClintock}, and
  {Richard}}}]{woods09}
{Woods}, T.~N., P.~C. {Chamberlin}, J.~W. {Harder}, R.~A. {Hock}, M.~{Snow},
  F.~G. {Eparvier}, J.~{Fontenla}, W.~E. {McClintock}, and E.~C. {Richard}
  (2009), {Solar Irradiance Reference Spectra (SIRS) for the 2008 Whole
  Heliosphere Interval (WHI)}, \textit{Geophys. Res. Lett.}, \textit{36},
  L1101, \doi{10.1029/2008GL036373}.

\bibitem[{\textit{{Yeo}}(2014)}]{yeo14c}
{Yeo}, K.~L. (2014), {Analysis and modeling of solar irradiance variations},
  Ph.D. thesis, Braunschweig University of Technology.

\bibitem[{\textit{{Yeo} et~al.}(2014{\natexlab{a}})\textit{{Yeo}, {Krivova},
  and {Solanki}}}]{yeo14a}
{Yeo}, K.~L., N.~A. {Krivova}, and S.~K. {Solanki} (2014{\natexlab{a}}), {Solar
  Cycle Variation in Solar Irradiance}, \textit{Space Sci. Rev.}, \textit{186},
  137--167, \doi{10.1007/s11214-014-0061-7}.

\bibitem[{\textit{{Yeo} et~al.}(2014{\natexlab{b}})\textit{{Yeo}, {Krivova},
  {Solanki}, and {Glassmeier}}}]{yeo14b}
{Yeo}, K.~L., N.~A. {Krivova}, S.~K. {Solanki}, and K.~H. {Glassmeier}
  (2014{\natexlab{b}}), {Reconstruction of total and spectral solar irradiance
  from 1974 to 2013 based on KPVT, SoHO/MDI, and SDO/HMI observations},
  \textit{Astron. Astrophys.}, \textit{570}, A85,
  \doi{10.1051/0004-6361/201423628}.

\end{thebibliography}

\end{article}

\begin{table}
\caption{Definition of the abbreviations featured in this paper.}
\label{acronyms}
\centering
\begin{tabularx}{\linewidth}{lX} 
\hline\hline
Abbreviation & Definition \\
\hline
ATLAS & ATmospheric Laboratory of Applications and Science \\
COSIR & COde for Solar Irradiance Reconstruction \\
HMI & Helioseismic and Magnetic Imager \\
LASP & Laboratory for Atmospheric and Space Physics \\
MDI & Michelson Doppler Imager \\
Mea11 & Morrill et al. (2011) \\
NOAA & National Oceanic and Atmospheric Administration \\
NRLSSI & Naval Research Laboratory Solar Spectral Irradiance \\
OAR & Observatorio Astronomico di Roma \\
PSI & Photometric Sunspot Index \\
SATIRE(-S) & Spectral And Total Irradiance REconstruction (for the Satellite era) \\
SBUV & Solar Backscatter Ultraviolet Radiometer \\
SDO & Solar Dynamics Observatory \\
SEE & Solar Extreme-ultraviolet Experiment \\
SIM & Spectral Irradiance Monitor \\
SME & Solar Mesosphere Explorer \\
SoHO & Solar and Heliospheric Observatory \\
SOLSPEC & SOLar SPECtrum \\
SOLSTICE & SOLar STellar Irradiance Comparison Experiment \\
SORCE & SOlar Radiation and Climate Experiment \\
SRPM & Solar Radiation Physical Modelling \\
SSI & Spectral Solar Irradiance \\
SUSIM & Solar Ultraviolet Spectral Irradiance Monitor \\
TIMED & Thermosphere Ionosphere Mesosphere Energetics and Dynamics \\
TSI & Total Solar Irradiance \\
UARS & Upper Atmosphere Research Satellite \\
UV & Ultraviolet \\
VIRGO & Variability of IRradiance and Gravity Oscillations \\
WHI & Whole Heliosphere Interval \\
\hline          
\end{tabularx}
\end{table}

\begin{sidewaystable}
\caption{Summary description of the SSI records examined in this study.}
\label{ssirecords}
\centering
\begin{tabular}{lccc} 
\hline\hline
SSI record (version) & Period [year.month.day] & Spectral range/resolution & Reference(s) \\
\hline
Nimbus-7/SBUV & 1978.11.08 to 1986.10.28 & 170 to 400 nm/1 nm & \cite{deland01} \\
SME & 1981.10.08 to 1989.04.12 & 115 to 303 nm/1 nm & \cite{rottman88} \\
NOAA-9/SBUV2 & 1985.03.14 to 1997.05.03 & 170 to 400 nm/1 nm & \cite{deland04} \\
NOAA-11/SBUV2 & 1988.12.05 to 1994.10.06 & 170 to 400 nm/1 nm & \cite{cebula98} \\
UARS/SUSIM (22) & 1991.10.12 to 2005.07.31 & 115 to 411 nm/1 nm & \cite{brueckner93,floyd03} \\
UARS/SOLSTICE (18) & 1991.10.17 to 2001.09.24 & 119 to 420 nm/1 nm & \cite{rottman01} \\
TIMED/SEE (11) & 2002.02.09 to 2015.05.09\tablenotemark{a} & 0 to 191 nm/1 nm & \cite{woods05} \\
SORCE/SIM (22) & 2003.04.14 to 2015.05.02 & 240 to 2416 nm/variable\tablenotemark{b} & \cite{harder05a,harder05b} \\
SORCE/SOLSTICE (14) & 2003.05.14 to 2015.05.02 & 115 to 310 nm/1 nm & \cite{mcclintock05,snow05a} \\
\hline          
\end{tabular}
\tablenotetext{a}{The TIMED/SEE record is extended with the latest measurements, with a three day lag, daily. The data set used here, which extends to 9th May 2015, was downloaded from lasp.colorado.edu/lisird/ on 12th May 2015.}
\tablenotetext{b}{Based on considerations related to the signal-to-noise ratio, SORCE/SIM measurements between 200 and 240 nm are not included in the present public release. The spectral resolution decreases with wavelength, from 0.01 nm at 240nm to 0.55 nm at 400 nm and eventually 11 nm at 2416 nm.}
\end{sidewaystable}

\begin{table}
\caption{Reported 1-sigma long-term uncertainty of the SSI records examined here.}
\label{reportedlongtermuncertainty}
\centering
\begin{tabular}{lcc} 
\hline\hline
SSI record (version) & Long-term uncertainty & Source \\
\hline
Nimbus-7/SBUV & 1.3$\%$ (205 nm) & \cite{deland01} \\
 & 0.9$\%$ (240 nm) & \\
 & 0.8$\%$ (300 nm) & \\
 & 0.8$\%$ (390 nm) & \\
SME & 1$\%$/year & \cite{rottman88} \\
NOAA-9/SBUV2 & 1.45$\%$ (180 nm) & \cite{deland04} \\
 & 1.15$\%$ (205 nm) & \\
 & 0.9$\%$ (250 nm) & \\
 & 0.6$\%$ (350 nm) & \\
NOAA-11/SBUV2 & 1.15$\%$ (180 nm) & \cite{cebula98} \\
 & 0.9$\%$ (205 nm) & \\
 & 0.9$\%$ (250 nm) & \\
 & 0.45$\%$ (350 nm) & \\
UARS/SUSIM (22) & 0.3$\%$/year (115 to 300 nm) & \cite{morrill14} \\
 & 0.03$\%$/year (300 to 411 nm) & \\
UARS/SOLSTICE (18) & 2$\%$ & \cite{rottman01} \\
TIMED/SEE (11) & 10$\%$ & \cite{woods05} \\
SORCE/SIM (22) & $<$0.1$\%$/year & Data header. \\
SORCE/SOLSTICE (14) & 0.2 to 0.5$\%$/year & Data header. \\
\hline          
\end{tabular}
\end{table}

\end{document}